\documentclass[aps,prb,twocolumn,superscriptaddress]{revtex4-1}
\usepackage[fleqn,tbtags]{amsmath}
\usepackage{amsfonts, amssymb,amsxtra,textcomp}
\usepackage[]{graphicx}
\usepackage{grffile}
\usepackage{bbm}
\usepackage{bm}
\usepackage{color}
\usepackage{graphicx}
\usepackage[colorlinks=true,citecolor=blue,linkcolor=blue]{hyperref}
\usepackage[utf8]{inputenc}

\newcommand{\bs}[1]{\boldsymbol{#1}}

\newcommand{\curO}{{\cal O}}

\newcommand{\Tr}{{\rm Tr}}

\newcommand{\be}{\begin{equation}}
\newcommand{\ee}{\end{equation}}
\newcommand{\bea}{\begin{eqnarray}}
\newcommand{\eea}{\end{eqnarray}}
\newcommand{\bse}{\begin{subequations}}
\newcommand{\ese}{\end{subequations}}

\usepackage{color}                                      
\definecolor{d_red}{cmyk}{0.00, 0.81, 1.00, 0.27}
\definecolor{d_orange}{cmyk}{0.00, 0.33, 1.00, 0.00}
\definecolor{d_blue}{cmyk}{0.78, 0.47, 0.00, 0.20}
\definecolor{d_lgreen}{cmyk}{0.07, 0.00, 0.79, 0.29}
\definecolor{d_green}{cmyk}{0.66, 0.00, 0.71, 0.56}
\definecolor{d_blue}{cmyk}{0.78, 0.47, 0.00, 0.20}
\definecolor{d_dblue}{cmyk}{0.91, 0.79, 0.00, 0.22}
\definecolor{d_pink}{cmyk}{0.0, 0.79, 0.37, 0.29}
\definecolor{d_purple}{cmyk}{0.16, 0.54, 0.00, 0.70}
\definecolor{d_paleblue}{cmyk}{0.669, 0.338, 0.00, 0.373}
\definecolor{d_dpaleblue}{cmyk}{0.441, 0.290, 0.00, 0.580}
\definecolor{d_brown}{cmyk}{0.0, 0.490, 0.930, 0.350}
\definecolor{d_turquoise}{cmyk}{0.630, 0.04, 0.0, 0.440}
\definecolor{KIT-green}{RGB}{0, 150,130}
\definecolor{KIT-blue}{RGB}{70,100,170}

\newcommand{\ii}{\mathrm{i}}

\newcommand{\im}{\text{Im}}

\def\bmx{\begin{pmatrix}}
\def\emx{\end{pmatrix}}

\usepackage{hyperref}
\usepackage[figure,table]{hypcap}               
\hypersetup{
pdftitle = {},
pdfsubject = {},
pdfauthor = {},
pdfkeywords = {},
pdfcreator = {},
pdfproducer = {LaTeX with hyperref},
colorlinks = true,
linkcolor = blue, 
anchorcolor = blue,
citecolor = blue, 
filecolor = red, 
menucolor = red, 
urlcolor  = blue, 
breaklinks = true,
pdfstartview = FitV,
pdfhighlight = /I,
pdfpagelayout = OneColumn,
hypertexnames=true 
}

\begin{document}

\title{Elastic response of the electron fluid in intrinsic graphene: the collisionless regime}

\author{Julia M. Link}
\affiliation{Institute for Theory of Condensed Matter, Karlsruhe Institute of
Technology (KIT), 76131 Karlsruhe, Germany}

\author{Daniel E. Sheehy}
\affiliation{Department of Physics and Astronomy, Louisiana State University,
Baton Rouge, LA, 70803, USA}

\author{Boris N. Narozhny }
\affiliation{Institute for Theory of Condensed Matter, Karlsruhe Institute of
Technology (KIT), 76131 Karlsruhe, Germany}
\affiliation{National Research Nuclear University MEPhI
(Moscow Engineering Physics Institute), 115409 Moscow, Russia}

\author{J\"org Schmalian}
\affiliation{Institute for Theory of Condensed Matter, Karlsruhe
  Institute of Technology (KIT), 76131 Karlsruhe, Germany}
\affiliation{Institute for Solid State Physics, Karlsruhe Institute of
  Technology (KIT), 76131 Karlsruhe, Germany}

\date{\today}
\begin{abstract} The elastic response of an electron fluid at 
finite frequencies is defined by the electron viscosity
$\eta(\omega)$.  We determine $\eta(\omega)$ for graphene at the
charge neutrality point in the collisionless regime, including the
leading corrections due to the electron-electron Coulomb
interaction. We find interaction corrections to $\eta(\omega)$ that
are significantly larger if compared to the 
corresponding corrections to the optical conductivity. In addition,
we find comparable contributions to the dynamic momentum flux due to
single-particle and many-particle effects. We also demonstrate that
$\eta(\omega)$ is directly related to the nonlocal energy-flow
response of graphene at the Dirac point.  The viscosity in the
collisionless regime is determined with the help of the strain
generators in the Kubo formalism. Here, the pseudo-spin of graphene
describing its two sublattices plays an important role in obtaining a
viscosity tensor that fulfills the symmetry properties of a
rotationally symmetric system.
\end{abstract}

\maketitle

\section{Introduction}

The low-frequency flow of electron charge and momentum in graphene is
dissipative \cite{Fritz2008,Kashuba2008,Mueller2009} and can be
described within the hydrodynamic approach
\cite{dau6,Forster,Lucas2018,Narozhny2017}. Signatures of hydrodynamic
behavior in graphene have been recently observed experimentally
\cite{Titov2013,Crossno2016,Ghahari2016,Bandurin,KrishnaKumar2017}
while attracting considerable theoretical attention
\cite{Briskot2015,Lucas2016,Levitov2016,Seo2017,Link2018}.

\begin{figure}
 \includegraphics[width=0.9\columnwidth]{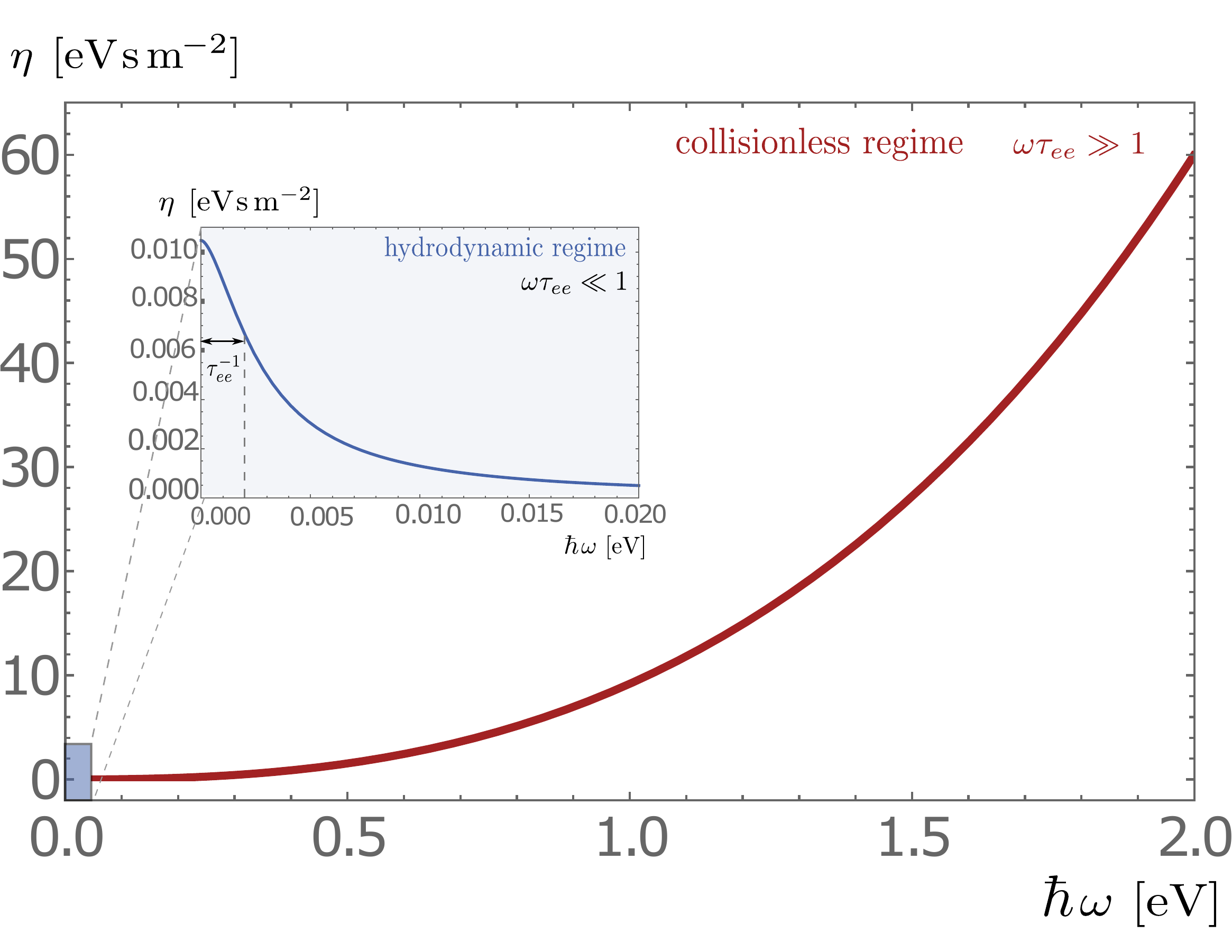}
 \caption{(Color Online) Main: Frequency-dependent shear viscosity,
   Eq.\eqref{eq:dynviscos}, in pure graphene at charge neutrality and
   temperature $T=0$ in the high-frequency collisionless
   regime. Inset: Frequency-dependent shear viscosity in the
   low-frequency hydrodynamic regime for $T=100$~K and an inverse
   scattering time of $\hbar \tau_{ee}^{-1}\simeq1.307 ~\alpha_0^2 k_B T
   \simeq0.002$~eV obtained by generalizing the analysis of
   Ref.~\onlinecite{Mueller2009} to small but finite
   frequencies~\cite{Kiselev2018}.}
 \label{fig:fig1} 
\end{figure}

The collective motion of charge carriers in a solid becomes
hydrodynamic, if the dominant scattering mechanism is provided by
electron-electron interactions, such that the corresponding scattering
rate $\tau_{ee}^{-1}$ multiplied by Planck's constant, $\hbar
\tau_{ee}^{-1}$, is the largest energy scale in the
problem~\cite{Narozhny2017,Lucas2018}. For example, the low-frequency, 
$\omega \tau_{ee} <1$, Drude-type dynamical viscosity due to  collisions 
of thermally excited carriers is shown in the inset of Fig.~\ref{fig:fig1}.
Stationary transport properties
are then encoded in a few kinetic coefficients describing dissipative
processes~\cite{Andreev2011}. In contrast to standard fluid
mechanics~\cite{dau6,Andreev2011}, dissipation in graphene is
described by the electrical (rather than the thermal)
conductivity at the charge neutrality point and the shear and bulk
viscosities~\cite{Lucas2018}. The former reflects the particular
property of Dirac fermions in graphene in which the energy current is
proportional to the total momentum and is conserved by
electron-electron interactions. The electric current is not conserved
and at charge neutrality can be relaxed by electron-electron
interactions \cite{Schuett2011,Narozhny2015}. The bulk viscosity in
graphene was argued to vanish
\cite{Mueller2009,Briskot2015,Narozhny2017}, at least within the
considered approximations.

At relatively high frequencies, i.e. in the optical collisionless
regime, $\omega\tau_{ee}\gg1$, free Dirac fermions in
  pure graphene at charge neutrality are characterized by the
  frequency-independent optical conductivity
  \cite{Sheehy2007,Herbut2008,Mishchenko2008,Sheehy2009,Golub2010,Juricic2010,Abedinpour2011,Sodemann2012,Rosenstein2013,Gazzola2013,Teber2014,Link2016},
  while electron-electron interactions yield a rather small, weakly
  frequency-dependent correction: 
\begin{equation}
  \sigma(\omega)
  =
  \frac{\pi e^{2}}{2h}\left[1+{\cal C}_{\sigma}\alpha(\omega)\dots\right],
  \quad
  \alpha(\omega)=\frac{\alpha_0}{1\!+\!\frac{\alpha_0}{4} \ln\frac{D}{\omega}}.
  \label{eq:conductivity}
\end{equation}
Here $\alpha(\omega)$ is a running (or renormalized) dimensionless
coupling constant measuring the strength of the Coulomb interaction,
$\alpha_0=e^{2}/(\hbar v\bar{\epsilon})$ is its bare value, $e$ is the
electron charge, $v$ is the bare velocity of the Dirac
fermions, $D$ is the bandwidth scale, and
${\bar{\epsilon}=(\epsilon_{1}\!+\!\epsilon_{2})/2}$ is determined by
the dielectric constants $\epsilon_{1,2}$ of the material above and
below the graphene sheet (in suspended graphene ${\bar{\epsilon}=1}$
and $\alpha_{0}\approx2.2$). Our results are obtaind using a perturbative 
renormalization group analysis. While $\alpha_0$ is of order unity, the expansion 
is in fact with regards to the renormalized coupling constant $\alpha(\omega)$ of 
Eq.~\ref{eq:conductivity} which is small for $\omega \ll D$.
The numerical coefficient
${\mathcal{C}_{\sigma}=(19-6\pi)/12\approx0.01}$ is rather small
\cite{Mishchenko2008,Sheehy2009,Teber2014,Link2016}. This is in
agreement with the experimental measurement \cite{Nair2008} of the
transmission coefficient related to the conductivity by
\cite{Stauber2008} ${T(\omega)=[1+2\pi\sigma(\omega)/c]^{-2}}$: the
measured value ${T\!=\!0.977}$ yields
${\sigma(\omega)\!\approx\!\pi{e}^{2}/(2h)}$.

In this paper we consider the electronic viscosity of graphene in the
collisionless regime, with a particular emphasis on understanding the
impact of electron-electron interactions on this observable. We start
by noting that the behavior of the electronic fluid in graphene at
finite frequencies is similar to that of an elastic medium
\cite{Conti1999}. Elastic deformations are described in terms of a
dynamic strain
\begin{equation}
\varepsilon_{\gamma\delta}(\bs{x},t)
=
\frac{\partial u_{\delta}(\bs{x},t)}{\partial x_{\gamma}},
\label{eq:strain}
\end{equation}
where the vector $\bs{u}(\bs{x},t)$ describes the displacement of a
fluid element. Within the elasticity theory, the strain
is linearly related to stress. If
$\langle\tau_{\alpha\beta}(\bs{x})\rangle$ is the expectation value of
the stress tensor operator and
$\delta\langle\tau_{\alpha\beta}\rangle$ is the change in $\langle
\tau_{\alpha\beta}\rangle$ from its value for zero strain,
$\varepsilon_{\gamma\delta}=0$, then we can define the dynamic 
elastic response constants by
\begin{equation}
\delta\langle\tau_{\alpha\beta}\rangle(t)
=
\int\limits_{-\infty}^{\infty}dt'C_{\alpha\beta\gamma\delta}(t-t')\varepsilon_{\gamma\delta}(t').
\label{eq:responseresult}
\end{equation}
In the collisionless regime, the dynamic elastic constants can be
split into isothermal and adsorptive parts (see
section~\ref{sec:viscoelastic_response_of_the_electron_fluid} for a
formal proof) with the latter related to the dynamic viscosity,
$\eta_{\alpha\beta\gamma\delta}(\omega)$:
\begin{equation}
C_{\alpha\beta\gamma\delta}(\omega)
=
C_{\alpha\beta\gamma\delta}^{T}-i\omega\eta_{\alpha\beta\gamma\delta}(\omega),
\label{eq:elastic}
\end{equation}
In an isotropic fluid, the elastic properties are determined by the
bulk modulus, hence the isothermal static elastic constant is equal to
\[
C_{\alpha\beta\gamma\delta}^{T}=\kappa_T^{-1}\delta_{\alpha\beta}\delta_{\gamma\delta},
\qquad
\kappa_T^{-1}=-V \left.\frac{\partial p}{\partial V}\right|_{N,T},
\]
with $\kappa_T$ the isothermal compressibility.

In systems with rotational invariance (which holds for graphene in the
low-energy limit), the tensor structure of the dynamic viscosity has
the form~\cite{dau6}
\begin{equation}
\label{eta}
  \eta_{\alpha \beta \gamma \delta}(\omega)
=
\eta(\omega)
\left[\delta_{\alpha \gamma} \delta_{\beta \delta}+ \delta_{\alpha \delta} \delta_{\beta \gamma}
-\frac{2}{d} \delta_{\alpha \beta} \delta_{\gamma \delta}\right],
\end{equation}
where $\eta(\omega)$ is the shear viscosity, $d$ is the
dimensionality, and we neglected the bulk viscosity
\cite{Mueller2009,Briskot2015,Narozhny2017}.

\begin{figure}
 \includegraphics[width=0.8\columnwidth]{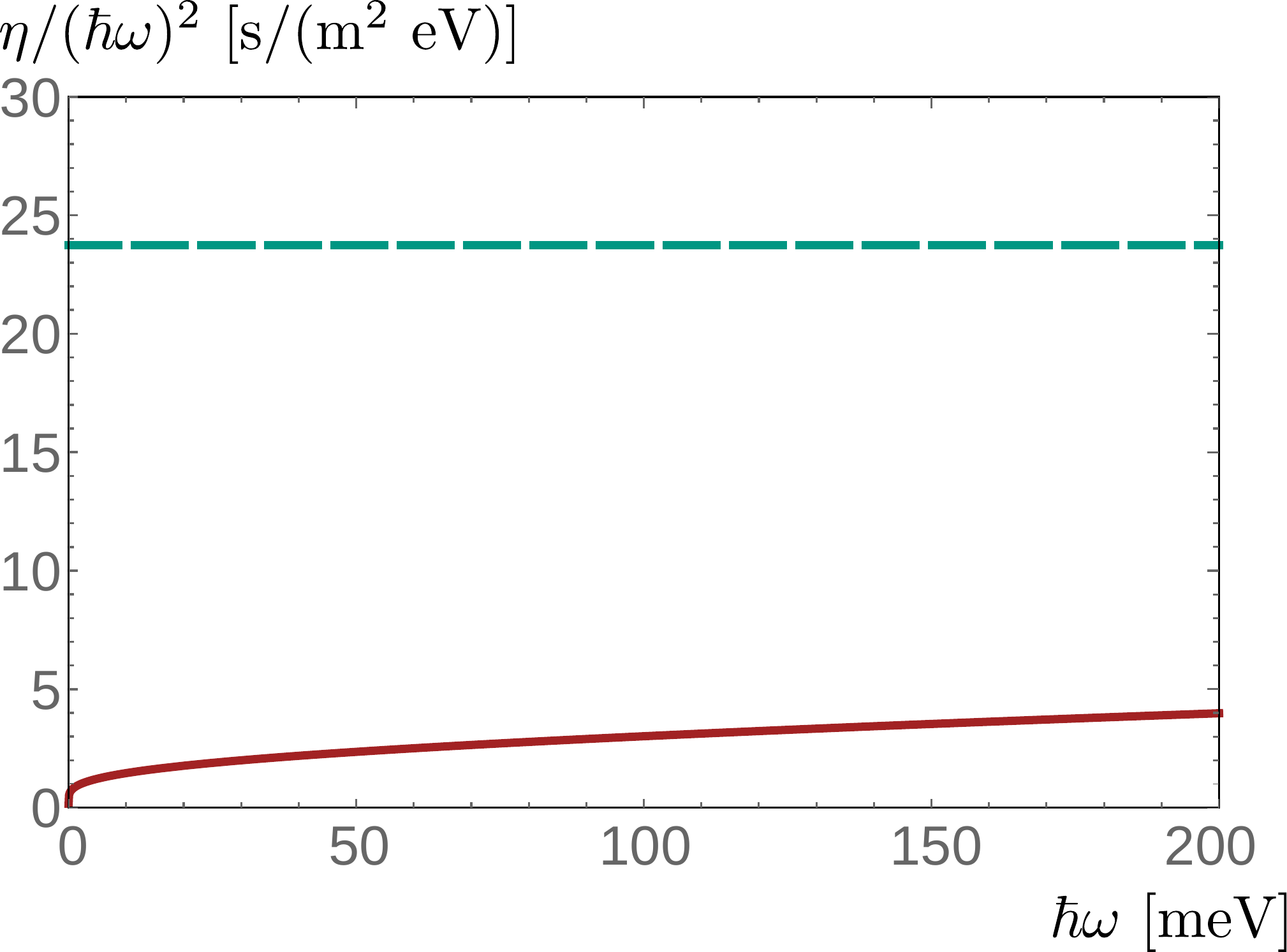}
 \caption{(Color Online) The effect of electron-electron interaction
   on the shear viscosity of pure graphene at charge neutrality in the
   high-frequency collisionless regime. The green, dashed line shows
   $\eta^{(0)}(\omega)/(\hbar \omega)^2$ for free Dirac fermions, see
   Eq.~(\ref{eq:visc-nonint}). The red curve shows our result
   (\ref{eq:dynviscos}) for the dynamical shear viscosity,
   $\eta(\omega)/(\hbar \omega)^2$, including the interaction effects also 
   shown in Fig.~\ref{fig:fig1}.}
 \label{fig:fig} 
\end{figure}

Our main result concerns $\eta(\omega)$ in the the collisionless
regime of pure graphene at the charge neutrality point (and at zero
temperature, $T=0$). We obtain
\begin{equation}
\eta(\omega)
=
\frac{\hbar}{64}
\frac{\omega^2}{v^2(\omega)}
\left[1+{\cal C}_{\eta}\alpha(\omega)+{\cal O}[\alpha^2(\omega)]\right],
\label{eq:dynviscos}
\end{equation}
with the numerical
coefficient
\[
{\cal C}_{\eta}=\frac{89-20\pi}{40}\approx0.65\gg C_\sigma.
\]
Here, $v(\omega)=v[1+(\alpha_0/4)\ln(D/\omega)]$ is the renormalized
velocity at frequency $\omega$.  The resulting frequency dependence of
the viscosity is shown in the main panel of Fig.~\ref{fig:fig1}, with
the inset illustrating the low-frequency hydrodynamic regime following
Refs.~\onlinecite{Mueller2009,Kiselev2018}. Our calculation for the 
collisionless regime is performed at zero temperature. Finite  $T$ will 
affect the low-frequency hydrodynamic regime shown in the inset of Fig.~\ref{fig:fig1}, 
while thermal effects are negligible for $\omega \gg k_{\rm B}T$. In Fig.~\ref{fig:fig} 
we compare $\eta(\omega)/\omega^2$ of  Eq.~\ref{eq:dynviscos} with and 
without Coulomb interactions, i.e. for $\alpha_0\neq 0$ and $\alpha_0=0$, 
respectively, in the collisionless regime. Coulomb corrections significantly 
suppress the viscosity in the regime $ \omega \ll D$.

Within the hydrodynamic theory of Galilean-invariant
  systems, dissipation affects the energy flow and the momentum flux
  as quantified by the thermal conductivity and the two viscosities,
  respectively \cite{dau6}. In particular, the shear viscosity,
  $\eta$, describes the tendency towards a uniform flow and can be
  directly
  related~\cite{Conti1999,Forcella2014,Principi2016,Bradlyn12} to the
  nonlocal optical conductivity, $\sigma(\bs{q},\omega)$. In contrast,
  the electronic hydrodynamics in graphene describes the energy flow
  \cite{Briskot2015} while dissipation affects the electric current
  (as well as the momentum flux). Here the shear viscosity determines
  the nonlocal thermal conductivity, $\kappa(\bs{q},\omega)$. At
  charge neutrality, the electric current and energy current are
  completely disentangled and orthogonal
  \cite{Briskot2015,Narozhny2015}, while the electrical conductivity
  is independent of viscosity
  \cite{Fritz2008,Lucas2018,Briskot2015,Narozhny2015}. The mutual
  independence of the electric and energy current manifests itself in
  the maximal violation of the Wiedemann-Franz law
  \cite{Crossno2016}. Far away from the neutrality point (i.e. for
  $|\mu|\ll T$), the electronic system in graphene behaves similarly
  to a Fermi liquid \cite{Lucas2018,Narozhny2017}. Here the energy
  current and the electric current are collinear and hence
  proportional to each other. Both currents now depend on both $\eta$
  \cite{Briskot2015,Principi2016} such that the above relation between
  $\sigma(\bs{q},\omega)$ and $\eta$ (as well as the Wiedemann-Franz
  law \cite{Crossno2016}) is restored. Finally, we note that the roles
  of frequency and temperature in this picture are not equivalent: the
  macroscopic currents can only be entagled by the collision
  integral. In the collisionless regime, the electric and energy
  currents remain orthogonal at charge neutrality;
  $\sigma(\bs{q},\omega)$ remains independent of $\eta$ even if
  $|\mu|\ll\omega$.

 Focusing on the charge neutrality point in graphene, we derive the
  relation between the viscosity and thermal conductivity using the
  Kubo formula approach of Ref.~\onlinecite{Bradlyn12} which we
  generalize to multi-component Dirac systems described by a
  pseudo-spin. This insight may be relevant not only to graphene, but
  also for other multi-band materials such as topological insulators,
  their surface states, Lieb lattices, and related systems.

Having briefly described our main results, we now describe the organization of the remainder of this
paper.  
In
section~\ref{sec:viscoelastic_response_of_the_electron_fluid} we
develop the formalism for the dynamic viscoelastic response of
graphene in the collisionless regime generalizing the framework of
Ref.~\onlinecite{Bradlyn12} to systems with a pseudo-spin structure,
including a formal proof of
Eq.~(\ref{eq:elastic}). Section~\ref{sec:viscosity} is devoted to the
calculation of the dynamic viscosity of pure graphene at charge
neutrality in the collisionless regime and it is split in two parts.
In the first part, Section~\ref{subsec:RG procedure},
we introduce the RG procedure which justifies our perturbative calculation
of the dynamic viscosity.
In the second part, 
Section~\ref{subsec:the case without Coulomb interaction}, we derive
the first term in Eq.\eqref{eq:dynviscos} which corresponds to the viscosity
of non-interacting Dirac fermions, while the interaction
correction is calculated in the following subsection. In
Section~\ref{sec:thermo} we derive the relation between
$\eta_{\alpha\beta\gamma\delta}(\omega)$ and $\kappa(\bs{q},\omega)$ and
in Section~\ref{conclusions} we provide concluding remarks.

\section{Viscoelastic response of the electron fluid}
\label{sec:viscoelastic_response_of_the_electron_fluid}

In this Section, we establish a general formalism for the dynamical
viscosity by generalizing the approach of Bradlyn \emph{et al.}~\cite{Bradlyn12}
to multi-band systems with a pseudo-spin structure. In graphene, the
pseudo-spin appears due to the sublattice structure of the honeycomb
lattice. The low-energy electronic excitations are described by a
spinor

\begin{equation}
  \label{spinor}
\psi(\bs{x})=
\begin{pmatrix}
\psi_{A}(\bs{x}) \cr
\psi_{B}(\bs{x})
\end{pmatrix} ,
\end{equation}
comprising the annihilation operators of electrons belonging to the
sublattice $A$ or $B$. In terms of the spinors, the effective
low-energy Hamiltonian of pure graphene at charge neutrality is
\begin{subequations}
\begin{equation}
H_{\Omega(t)} =
H_{0,\Omega(t)}+H_{\mathrm{int},\Omega(t)}\:,
\end{equation}
\begin{equation}
H_{0,\Omega(t)}
=
v\!\int\limits_{\Omega(t)}\!d^{d}x
\psi^{\dagger}(\bs{x}) \; \bs{\sigma}\!\cdot\!\hat{\bs{p}} \; \psi(\bs{x})
\:,
\label{eq:H0 graphene}
\end{equation}
\begin{equation}
H_{\mathrm{int},\Omega(t)}
=
\frac{1}{2}\!\int\limits_{\Omega(t)}\!d^{d}xd^{d}y
\psi^{\dagger}(\bs{x})\psi^{\dagger}(\bs{y})
V(\bs{x}-\bs{y})
\psi(\bs{y})\psi(\bs{x}),
\end{equation}
\end{subequations}
where $\Omega\left(t\right)$ describes a time dependent domain within
which the electronic systems is assumed to be confined 
and the dimensionality $d$ is given by $d=2$ for graphene. For simplicity
we also use $\Omega\left(t\right)$ to denote the volume of this
domain. Here $v$ denotes the bare velocity, $\hat{\bs{p}}$ is the
momentum operator, and $\bs{\sigma}$ is the vector of the Pauli
matrices acting in the pseudo-spin space. The additional valley and
spin degrees of freedom give rise to an overall degeneracy factor
$N=4$ in the viscosity. The electrons interact by the Coulomb law,
$V(\bs{x}-\bs{y})=e^{2}/(4\pi\bar{\epsilon}|\bs{x}-\bs{y}|)$.

\subsection{Strain generators for systems with pseudo-spin }

Following Ref.~\onlinecite{Bradlyn12}, we analyze generic coordinate
transformations, $\bs{x}\rightarrow\bs{x}'=\bs{x}+\bs{u}(\bs{x},t)$,
which are realized in terms of a homogeneous but time-dependent
invertible ${d\times{d}}$ matrix $\widehat{\Lambda}(t)$ with a positive
determinant via
\begin{equation}
\bs{x}'=\widehat{\Lambda}(t)^{T}\bs{x}.
\label{eq:transformation}
\end{equation}
The matrix $\widehat{\Lambda}(t)$ can be expressed in terms of the
strain tensor $\hat{\varepsilon}(t)$ of Eq.~\eqref{eq:strain}:
\begin{equation}
  \label{lambda}
\widehat{\Lambda}(t)=e^{\hat{\varepsilon}(t)}.
\end{equation}
Indeed, for small strain we have 
$\bs{x}'\approx(\hat{1}+\hat{\varepsilon}^{T})\bs{x}$ with the
displacement $\bs{u}(\bs{x})=\hat{\varepsilon}^{T}\bs{x}$ and we
recover Eq.~(\ref{eq:strain}).

Usually, the viscous dynamics is expressed in terms of the response to
a velocity gradient. This follows from the relation
\begin{equation}
\frac{\partial v_{\delta}}{\partial x_{\gamma}}=\frac{\partial\varepsilon_{\gamma\delta}}{\partial t}.
\end{equation}
The coordinates $x'_{\alpha}$ can be strained due to rotations, shear,
or compressions. All transformations should adiabatically connect to
the unit matrix, which implies the abovementioned restriction ${\rm
  det}\widehat{\Lambda}>0$.

\subsubsection{Uniform compressions}

In order to illustrate the more general case of the next paragraph we
first consider uniform compressions which represent a physical system with a
time-dependent volume $\Omega\left(t\right)$ that preserves its shape
and orientation. Our goal is to express the dynamics of
$H_{\Omega\left(t\right)}$ in terms of a Hamiltonian with a fixed
volume $\Omega_{0}$ and additional perturbations. For homogeneous
compressions we write for the transformation matrix:
\begin{equation}
\widehat{\Lambda}(t)=e^{\varepsilon\left(t\right)}\hat{1}
\:,
\end{equation}
where the strain tensor is diagonal,
${\hat{\varepsilon}=\varepsilon\hat{1}}$. The trace of the strain tensor,
${{\rm Tr}\;\hat{\varepsilon}=d\varepsilon}$, determines the relative
volume change, $\Omega\left(t\right)=e^{d\varepsilon(t)}\Omega_{0}$
(in the case of graphene $d=2$).

By requiring that canonical anticommutation relations are preserved, we find
the form of the electron field operator after the transformation
\begin{equation}
\psi_{\varepsilon}(\bs{x})=e^{d\varepsilon(t)/2}\psi\left(e^{\varepsilon(t)}\bs{x}\right).
\end{equation}
The same transformation can be expressed in terms of a unitary operator
\begin{equation}
  \psi_{\varepsilon}(\bs{x})=U(t)\psi(\bs{x}),
  \qquad
  U(t)=e^{-i\varepsilon(t){\rm Tr}\hat{{\cal L}}}.
  \label{eq:unitary transf}
\end{equation}
The infinitesimal generator $\hat{{\cal L}}$ was introduced in
Ref.~\onlinecite{Bradlyn12} as the ``strain generator''. The explicit
form of the strain generator can be inferred from the requirement that
the two forms of the transformation yield identical results
\cite{Bradlyn12}. For uniform compressions, the strain generator is
diagonal and is given by
\begin{equation}
{\cal L}_{\alpha\alpha}=\frac{i}{2}+ix_{\alpha}\frac{\partial}{\partial x_{\alpha}}.
\label{EQ:strain gen Laa}
\end{equation}

Since Eq.~\eqref{eq:unitary transf} describes a time-dependent
transformation, we obtain the additional contribution to the
Hamiltonian
\begin{equation*}
  \delta H_{t}=-i\!\int\! d^{d}x\psi^{\dagger}(\bs{x},t)U(t)
  \left[\partial_{t}U^{-1}(t)\right]\psi(\bs{x},t).
\end{equation*}
Finally, for arbitrary functions of the momentum or position
operators, $f(\bs{p})$ or $g(\bs{x})$, respectively, it holds 
\begin{eqnarray*}
  &&
  U^{-1}f\left(e^{-\varepsilon(t)}\bs{p}\right)U\!=\!f(\bs{p}),
  \\
  &&
  \\
  &&
  U^{-1}g\left(e^{\varepsilon(t)}\bs{x}\right)U\!=\!g(\bs{x}).
\end{eqnarray*}
Combining the above expressions, we express the Hamiltonian
of a system with time-dependent volume as 
\begin{equation}
  H_{\Omega(t)}=H_{\Omega_{0}}
  -
  \!\int\! d^{d}x\psi^{\dagger}(\bs{x})\frac{\partial\varepsilon}{\partial t}
  \sum_{\alpha}{\cal L}_{\alpha\alpha}\psi(\bs{x}),
\end{equation}
where $H_{\Omega_{0}}$ is the Hamiltonian with a fixed volume
$\Omega_0$. Thus, we see that the time-dependent compression can be ``gauged away'' and
yields a form of the Hamiltonian that can be treated within the
usual Kubo formalism, see section~\ref{kubo}.

We note that the internal spinor structure of the fermion field did not play any
role in the above analysis of uniform compressions. However, in our  subsequent
discussion of arbitrary dynamical strain this will no longer be the
case.

\subsubsection{Arbitrary dynamical strain}

Now we study arbitrary strain fields, such that $\hat{\Lambda}(t)$ in
Eq.~\eqref{eq:transformation} is an arbitrary matrix with positive
determinant. Following Ref.~\onlinecite{Bradlyn12}, the transformation
of the field operators takes the form
\begin{equation}
\psi_{\varepsilon}(\bs{x})=\sqrt{\det\Lambda}\psi\left(\Lambda^{T}\bs{x}\right),
\end{equation}
where the factor $\sqrt{\det\Lambda}$ ensures the proper canonical
commutation relation (i.e., normalization).

For an infinitesimal change from $\hat{\varepsilon}$ to
$\hat{\varepsilon}+\delta \hat{\varepsilon}$ we find
\begin{equation}
  \frac{\partial}{\partial\varepsilon_{\alpha\beta}}\psi_{\varepsilon}(\bs{x})
  =\frac{\delta_{\alpha\beta}}{2}\psi_{\varepsilon}(\bs{x})
  +x'_{\alpha}\frac{\partial}{\partial x'_{\beta}}\psi_{\varepsilon}(\bs{x}).
\end{equation}
Here the first term stems from the change in the determinant, while
the second term is the derivative with respect to the coordinates
$\bs{x}'$. For infinitesimal changes we may set $\hat{\varepsilon}=0$ on the right side
so 
that there is no distinction between the two sets of coordinates in the
second term. As a result, the infinitesimal strain transformation can
be expressed as
\begin{eqnarray}
  \psi_{\varepsilon}(\bs{x})
  & = &
  \psi(\bs{x})+\sum_{\alpha\beta}\varepsilon_{\alpha\beta}
  \left.\frac{\partial\psi_{\varepsilon}(\bs{x})}{\partial\varepsilon_{\alpha\beta}}\right|_{\varepsilon=0}
  \:,
  \nonumber\\
  &&
  \nonumber\\
  & = &
  \left[1-i\sum_{\alpha\beta}\varepsilon_{\alpha\beta}{\cal L}_{\alpha\beta}\right]\psi(\bs{x}),
\end{eqnarray}
with 
\begin{equation}
  {\cal L}_{\alpha\beta}
  =
  ix_{\alpha}\frac{\partial}{\partial x_{\beta}}
  +\frac{i}{2}\delta_{\alpha\beta}
  = - \frac{1}{2}\left\{ x_{\alpha},p_{\beta}\right\},
  \label{eq: Lab}
\end{equation}
playing the role of the generators of this transformation and generalizing Eq.~(\ref{EQ:strain gen Laa}).

Extending these arguments to finite displacements \cite{Bradlyn12}, we
arrive at the natural generalization of Eq.~\eqref{eq:unitary transf}
\begin{equation}
\psi_{\varepsilon}(\bs{x})=U_{{\cal L}}(t)\psi(\bs{x}),
\quad
U_{{\cal L}}(t)=e^{-i{\rm Tr}\left(\hat{\varepsilon}^{T}\hat{{\cal L}}\right)}.
\end{equation}
The form of the strain generators (\ref{eq: Lab}) can also be
established \cite{Bradlyn12} by considering the transformations of the
coordinate and momentum operators,
\[
\Lambda^{T}\bs{x} = U_{{\cal L}} \bs{x} U_{{\cal L}}^{-1},
\quad
\Lambda^{-1}\bs{p} = U_{{\cal L}} \bs{p} U_{{\cal L}}^{-1},
\]
which lead to the commutation relations
\[
  [{\cal L}_{\mu\nu}, x_\alpha] = i \delta_{\alpha\nu} x_\mu,
  \quad
  [{\cal L}_{\mu\nu}, p_\alpha] = -i \delta_{\alpha\mu} p_\nu,
\]
satisfied by Eq.~(\ref{eq: Lab}). The resulting transformation of the
momentum operator is
\begin{equation}
U_{{\cal L}}p_{\alpha}U_{{\cal L}}^{-1}=p_{\alpha}+\varepsilon_{\alpha\beta}p_{\beta}.
\end{equation}

We now
recall that the generic coordinate transformations
(\ref{eq:transformation}) also include spatial rotations. This implies that
 the infinitesimal generators (\ref{eq: Lab}) are related
to the angular momentum. For fermion fields without internal degrees
of freedom, the usual (orbital) angular momentum is determined by the
antisymmetric part of the strain generator \cite{Bradlyn12},
${L_{\alpha}=-\varepsilon_{\alpha\beta\gamma}{\cal
    L}_{\beta\gamma}=\varepsilon_{\alpha\beta\gamma} x_\beta p_\gamma}$.
However, as is well-known from standard field theory \cite{dau4},
the proper generators of  infinitesimal rotations are the operators
of total angular momentum. In the case of graphene,
two-dimensional in-plane rotations are generated by the component of
the total angular momentum orthogonal to the graphene sheet
\cite{Mecklenburg2011}. Since both sublattices are affected by the
rotations, the total angular momentum of the Dirac fermions in
graphene includes the pseudo-spin.

A natural form of the Hermitian operator that corresponds to strain
transformations in pseudo-spin space is
\begin{equation}
  {\cal S}_{\alpha\beta}=\frac{i}{8}\left[\sigma_{\alpha},\sigma_{\beta}\right]
  =-\frac{1}{4}\varepsilon_{\alpha\beta\gamma}\sigma_{\gamma},
\end{equation}
which yields the desired relation to the pseudo-spin contribution to
the angular momentum,
${S_{\alpha}=-\varepsilon_{\alpha\beta\gamma}{\cal{S}}_{\beta\gamma}}$. The
tensor ${\cal S}_{\alpha\beta}$ is asymmetric. The only possible
choice for a symmetric contribution would be proportional to
$\{\sigma_{\alpha},\sigma_{\beta}\}=2\delta_{\alpha\beta}\sigma_{0}$,
which is trivial in pseudo-spin space. For the unitary
transformation in pseudo-spin space we find therefore
\[
  U_{{\cal S}}=e^{-i{\rm Tr\left(\hat{\varepsilon}{\cal \hat{S}}\right)}}
  \quad\Rightarrow\quad e^{i\left(\varepsilon_{xy}-\varepsilon_{yx}\right)\sigma_{z}/2} \:,
\]
where the latter expression is specific for $d=2$. 

The preceding arguments show that the correct
strain generator of graphene is
\begin{equation}
{\cal J}_{\alpha\beta} = {\cal L}_{\alpha\beta}+{\cal S}_{\alpha\beta}
= -\frac{1}{2}\left\{ x_{\alpha},p_{\beta}\right\}
+\frac{i}{8}\left[\sigma_{\alpha},\sigma_{\beta}\right],
\label{eq:stress generator J}
\end{equation}
such that the transformation matrix is given by
\begin{equation}
U = U_{{\cal S}}U_{{\cal L}}=e^{-i{\rm Tr\left(\hat{\varepsilon}{\cal \hat{J}}\right)}}.
\end{equation}

Generalizing the above arguments for the case of uniform compressions,
we arrive at the following form of the Hamiltonian in a general
time-dependent domain, $\Omega\left(t\right)$,
\begin{equation}
  H_{\Omega\left(t\right)}
  =H_{\Omega_{0}}-\!\int\! d^{d}x
  \sum_{\alpha\beta}\frac{\partial\varepsilon_{\alpha\beta}}{\partial t}
  \psi^{\dagger}(\bs{x}){\cal J}_{\alpha\beta}\psi(\bs{x}).
  \label{eq:coupling Hamiltonian}
\end{equation}
Thus, a time dependent strain field couples to the strain generator
${\cal J}_{\alpha\beta}$ that affects the coordinates and
pseudo-spin structure. This result will enable us to determine the
proper (symmetric) stress tensor and the Kubo formula for the
viscosity of the Dirac fermions in graphene.

\subsection{Momentum conservation and stress tensor}

Having determined the form of graphene's strain tensor
$\varepsilon_{\beta\alpha}(\bs{x},t)$ and its coupling to the electron
fluid, taking into account the sublattice structure of the honeycomb
lattice, our next task is to identify the stress tensor
$\tau_{\beta\alpha}(\bs{x},t)$. The linear-response relationship
between these tensors, given above in Eq.~(\ref{eq:responseresult}),
then defines the viscosity tensor.

To obtain the stress tensor, we begin by recalling that, 
in a translationally invariant system, momentum is conserved and the
operators 
\begin{equation}
  G_{\alpha}=-i\!\int\! d^{d}x\;
  \psi^{\dagger}(\bs{x},t)\partial_{\alpha}\psi(\bs{x},t) 
  \label{eq:momentum}
\end{equation}
of the $\alpha$-th component of the total momentum commute with the
Hamiltonian. Momentum conservation can also be expressed by the
continuity equation
\begin{equation}
\label{Eq:continuity}
\partial_{t}g_{\alpha}(\bs{x},t)+\partial_{\beta}\tau_{\beta\alpha}(\bs{x},t)=0,
\end{equation}
where $g_{\alpha}(\bs{x},t)$ is the momentum density
\begin{equation}
  g_{\alpha}(\bs{x},t)=-i\psi^{\dagger}(\bs{x},t)\partial_{\alpha}\psi(\bs{x},t),
  \label{eq:original g}
\end{equation}
and $\tau_{\alpha\beta}(\bs{x},t)$ is the momentum flux or stress
tensor.

The choice (\ref{eq:original g}) of the momentum density is, however,
not unique \cite{dau2,book}: adding a contribution acting as a surface
term in the integration of Eq.~\eqref{eq:momentum} does not change the
total momentum. In the standard field theory \cite{dau2,book} this
freedom is used to bring the canonical stress tensor to a symmetric
form that is typically assumed in calculations of the viscosity tensor
\cite{Principi2016} (using an asymmetric stress tensor leads to
results that are explicitly incorrect).

In what follows, we modify the momentum density (\ref{eq:original g})
(preserving the total momentum) to \cite{book}
\begin{eqnarray}
  &&
  g_{\alpha}(\bs{x},t)
  =
  \frac{i}{4}\left\{
    \left[\partial_{\alpha}\psi^{\dagger}(\bs{x},t)
      +\bs{\nabla}\psi^{\dagger}(\bs{x},t)\!\cdot\!\bs{\sigma}\sigma_\alpha\right]\psi(\bs{x},t)
    \right.
  \nonumber\\
  &&
  \nonumber\\
  &&
  \qquad  
  -
  \left.
  \psi^{\dagger}(\bs{x},t)
  \left[\partial_{\alpha}\psi(\bs{x},t)+\sigma_\alpha\bs{\nabla}\psi(\bs{x},t)\!\cdot\!\bs{\sigma}\right]
  \right\},
 \label{eq:modified g}
\end{eqnarray}
in order to derive a symmetric form of
$\tau_{\alpha\beta}(\bs{x},t)$. The latter can be found by considering
the long-wavelength limit of the continuity equation
(\ref{Eq:continuity}). Indeed, applying a Fourier transformation
with respect to the spatial coordinates we may write
Eq.~(\ref{Eq:continuity}) in the form
\[
\partial_{t}g_{\alpha}(\bs{q},t)-iq_{\beta}\tau_{\beta\alpha}(\bs{q},t)=0.
\]
Expanding the Fourier-transformed momentum density 
$g_{\alpha}(\bs{q},t)$ for small $\bs{q}$, we find
\begin{eqnarray}
  g_{\alpha}(\bs{q},t)
  & = &
  \int\! d^{d}x \; e^{i\bs{q}\cdot\bs{x}}g_{\alpha}\left(\bs{x},t\right) \:,
  \\
  &&
  \nonumber\\
  & \approx &
  g_{\alpha}\left(0,t\right)+iq_{\beta}\!\int\! d^{d}x\;\psi^{\dagger}\left(\bs{x},t\right)
  \nonumber\\
  &&
  \nonumber\\
  & \times &
  \left[x_{\beta}\left(-i\frac{\partial}{\partial x_{\alpha}}\right)
    +\frac{1}{4}\epsilon_{\beta\alpha\gamma}\sigma_{\gamma}\right]\psi\left(\bs{x},t\right)+\dots.
  \nonumber
\end{eqnarray}
To leading order in small $\bs{q}$, this formula implies conservation of the total momentum 
$\partial_{t}g_{\alpha}\left(\bs{0},t\right)=\partial_{t}G_{\alpha}\left(t\right)=0$, while 
the first subleading order reveals
\begin{equation}
  \partial_{t}{\cal J}_{\alpha\beta}=-T_{\alpha\beta},
  \label{contin}
\end{equation}
where
\[
T_{\alpha\beta}=\tau_{\alpha\beta}\left(\bs{q}=\bs{0}\right)=\!\int\! d^dx \; \tau_{\alpha\beta}(\bs{x}),
\]
is the integrated stress tensor and ${\cal J}_{\alpha\beta}$ is the
stress generator of Eq.~\eqref{eq:stress generator J}. As a result,
the commutator of the stress generator with the Hamiltonian yields the
explicitly symmetric stress tensor
\begin{equation}
  T_{\alpha\beta}=-i\left[H,{\cal J}_{\alpha\beta}\right].
  \label{eq:contin2}
\end{equation}
We postpone evaluating this commutator until after we obtain the Kubo formula expression for
graphene's viscosity.  Before proceeding to this task we note that 
in the case of a rotationally invariant system, the resulting stress
tensor is equivalent to the Belinfante-Rosenfeld stress-energy tensor
in the usual Dirac theory \cite{Belinfante1940,Rosenfeld1940}. 
The approach presented here, however, has the advantage that it may also be applied to anisotropic systems such as 
the anisotropic Dirac fluids studied in Ref.~\onlinecite{Link2018}.

\subsection{Kubo formalism for the viscosity tensor}
\label{kubo}

We now proceed with the development of the Kubo formalism for the dynamic
viscosity of graphene. To make our presentation self-contained, we begin by
summarizing the usual linear response theory
\cite{Kubo1957,Mahan}. Consider a system subjected to an external,
time-dependent perturbation
\begin{equation}
\delta H=-\sum_{j}A_{j}F_{j}\left(t\right),\label{eq: pert}
\end{equation}
characterized by the operators $A_{j}$ and time dependent functions
$F_{j}\left(t\right)$. Within linear response, the expectation values
$\langle A_{i}\rangle_{t}$ acquire an additional contribution  
\begin{equation}
  \delta\langle A_{i}\rangle_{t}
  =
  \!\int\limits_{-\infty}^{\infty}\!dt\sum_{j}G_{ij}(t-t')F_{j}(t'),
\end{equation}
where
\[
G_{ij}(t)=-i\theta(t)\langle[A_{i}(t),A_{j}(0)]\rangle,
\]
is the retarded Green's function. In addition, one may make use of the
Kubo identity
\[
i\left[A(t),\rho\right]=\rho\!\int\limits_{0}^{\beta}\!d\tau\dot{A}(t-i\tau),
\]
perform a Fourier transformation and partial integration, and obtain
\begin{equation}
  G_{ij}(\omega)=\frac{i}{\omega+i0^{+}}\left[\chi_{ij}(\omega)-\chi_{ij}^{T}\right],
  \label{eq:Kubo2}
\end{equation}
where
\[
\chi_{ij}(t-t')=-i\theta(t-t')
\left\langle\left[A_{i}(t),\dot{A}_{j}(t')\right]\right\rangle,
\]
and
\[
\chi_{ij}^{T}=\left.
\frac{\partial\langle A_{i}\rangle}{\partial F_{j}^{{\rm stat}}}\right|_{F_{j}^{{\rm stat}}=0},
\]
is the isothermal susceptibility due to an external static field
$F_{j}^{{\rm stat}}$ coupling to $\dot{A}_{j}$ in the
Hamiltonian \cite{Kubo1957}.

Now we apply the above Kubo formalism to the viscosity tensor defined
by the linear response relation
\begin{equation}
  \delta\langle\tau_{\alpha\beta}\rangle_{t}
  \!=\!
  \langle\tau_{\alpha\beta}\rangle_{\bs{x}'}-\langle\tau_{\alpha\beta}\rangle_{\bs{x}}
  \!=\!
  \!\!\int\limits_{-\infty}^{\infty}\!\!\!dt
  \sum_{\gamma\delta}\eta_{\alpha\beta\gamma\delta}(t\!-\!t')
  \frac{\partial\varepsilon_{\gamma\delta}}{\partial t},
\label{eq:Kubo-viscosity-1}
\end{equation}
where $\langle\tau_{\alpha\beta}\rangle_{\bs{x}'(\bs{x})}$ indicates
the stress tensor averaged over the system with the deformed or
undeformed coordinates, respectively.

The time dependent strain field $\varepsilon_{\alpha\beta}(t)$ couples to
the system by means of Eq.~\eqref{eq:coupling Hamiltonian}. At the
same time, the integrated stress tensor is proportional to the time
derivative of ${\cal J}_{\alpha\beta}$, see Eqs.~\eqref{contin},
\eqref{eq:contin2}. Here one has to distinguish between the integrated
stress tensor $T_{\alpha \beta}$ and the local stress tensor
$\tau_{\alpha \beta}(\bs{q}=0)$ which are connected via
\begin{equation}
  \langle T_{\alpha\beta}\rangle
  =\!\int\limits_{\Omega}\!d^d x\langle\tau_{\alpha\beta}(\bs{x})\rangle
  =V\langle\tau_{\alpha\beta}(\bs{q}=0)\rangle.
\label{eq:relation-stress-integratedstress}
\end{equation}
In order to determine the viscosity, we need to calculate
$\langle\tau_{\alpha\beta}\rangle_{\bs{x}'}$, i.e., the expectation
value of the stress tensor in the deformed coordinate system, see
Eq.~\eqref{eq:Kubo-viscosity-1}. Using the perturbation
defined in Eq.~\eqref{eq:coupling Hamiltonian} we may determine the
expectation value of the integrated stress tensor
\begin{equation}
\langle T_{\alpha\beta}\rangle_{\bs{x}'}
=\langle T_{\alpha\beta}\rangle_{\bs{x}}+
\!\int\limits_{-\infty}^{\infty}\!dt\;
X_{\alpha\beta\gamma\delta}\,\frac{\partial\varepsilon_{\gamma\delta}}{\partial t}
\:,
\label{eq:integrate-stress-deformed}
\end{equation}
where the correlation function $X_{\alpha\beta\gamma\delta}(\omega)$
is defined as
\begin{equation}
X_{\alpha\beta\gamma\delta}(\omega)
=
\frac{-1}{i(\omega\!+\!i0^{+})}
\left(\frac{\partial\left\langle T_{\alpha\beta}\right\rangle }{\partial\varepsilon_{\gamma\delta}}
+C_{\alpha \beta \gamma \delta}(\omega)\right)_{\bs{x}}\!,
\end{equation}
with $C_{\alpha \beta \gamma \delta}(\omega)$ being the 
Fourier transform of the stress-stress correlation
function 
\begin{equation}
  C_{\alpha\beta\gamma\delta}(t-t')=i\theta(t-t')
  \langle \left[T_{\alpha\beta}(t),T_{\gamma\delta}(t')\right]\rangle,
\label{eq:correl function}
\end{equation}
previously introduced in Eq.~\eqref{eq:responseresult}.

The correlation function $X_{\alpha\beta\gamma\delta}(\omega)$ can be
related to the local stress tensor using
Eq.~\eqref{eq:relation-stress-integratedstress} and the transformation
law $V_{\bs{x}'}=V_{\bs{x}}\exp({\rm Tr}\;\hat{\varepsilon})$,
\begin{equation}
  \langle\tau_{\alpha\beta}\rangle_{\bs{x}'}
  \!=\!
  (1-\delta_{\gamma\delta}\varepsilon_{\gamma\delta})\langle\tau_{\alpha\beta}\rangle_{\bs{x}}
  +
  \frac{1}{V_{\bs{x}'}}\!\!\int\limits_{-\infty}^{\infty}\!\!dt\;
  \left(X_{\alpha\beta\gamma\delta}\right)_{\bs{x}}\!\frac{\partial\varepsilon_{\gamma\delta}}{\partial t}.
\end{equation}
Hence, the viscosity tensor has the form
\begin{equation}
  \eta_{\alpha\beta\gamma\delta}(t)
  =
  X_{\alpha\beta\gamma\delta}(t)-\delta_{\delta\gamma}\langle\tau_{\alpha\beta}\rangle_{\bs{x}}\theta(t).
\end{equation}
Now we use the identity
\[
\varepsilon_{\gamma\delta}(t)
\!=\!
\int\limits_{-\infty}^{t}\!dt'\;\frac{\partial\varepsilon_{\gamma\delta}}{\partial t'}
=
\!\int\limits_{-\infty}^{\infty}\!dt'\;\theta(t)\frac{\partial\varepsilon_{\gamma\delta}}{\partial t'},
\]
to perform the Fourier transformation of the viscosity tensor
\begin{eqnarray}
  \eta_{\alpha\beta\gamma\delta}(\omega)
  &=&
  X_{\alpha\beta\gamma\delta}(\omega)
  -\frac{\langle\tau_{\alpha\beta}\rangle_{\bs{x}}\delta_{\gamma\delta}}{i(\omega\!+\!i0^{+})}
  \:,
  \nonumber\\
  &&
  \nonumber\\
  &=&
  X_{\alpha\beta\gamma\delta}(\omega)-\frac{\delta_{\gamma\delta}P}{i(\omega\!+\!i0^{+})},
\end{eqnarray}
where we have used the fact that the averaged stress tensor defines
the pressure of the system
\[
\langle\tau_{\alpha\beta}\rangle_{\bs{x}}=P\,\delta_{\alpha\beta}.
\]

Finally, using Eq.~\eqref{eq:Kubo2} we express
the dynamic viscosity as
\begin{equation}
  \eta_{\alpha\beta\gamma\delta}\left(\omega\right)
  =\frac{C_{\alpha\beta\gamma\delta}^{T}-C_{\alpha\beta\gamma\delta}(\omega)}{i(\omega\!+\!i0^{+})},
\label{eq:viscosity-tensor}
\end{equation}
where
\[
C_{\alpha\beta\gamma\delta}^{T}
=
-\left.\frac{d\langle\tau_{\alpha\beta}\rangle}
{d\varepsilon_{\gamma\delta}^{{\rm stat}}}\right|_{\varepsilon_{\alpha\beta}=0},
\]
is the isothermal elastic constant (we have added a term
$-T_{\gamma\delta}\varepsilon_{\gamma\delta}^{{\rm stat}}$ with static
strain $\varepsilon_{\gamma\delta}^{{\rm stat}}$ to the Hamiltonian). The
above argument constitutes a formal proof of Eq.~\eqref{eq:elastic}.

\section{Dynamic viscosity of graphene}
\label{sec:viscosity}

In this Section, we use the Kubo formula
Eq.~\eqref{eq:viscosity-tensor} to evaluate the dynamic viscosity
tensor of pure graphene at charge neutrality and in the collisionless
regime. Since we work at finite frequencies, we can drop the
delta-function part of Eq.~(\ref{eq:viscosity-tensor}) to arrive at
\begin{equation}
\label{eq:arriveat}
\eta_{\alpha \beta \gamma \delta}(\omega)=\frac{{\rm Im}\;C_{\alpha \beta \gamma \delta}(\omega)}{\omega}.
\end{equation}
Thus, we only need to compute the Fourier transform of the correlation
function \eqref{eq:correl function}.  In doing this, we shall combine
perturbation theory with the renormalization group (RG) in order to
arrive at the result Eq.~(\ref{eq:dynviscos}).

\subsection{RG procedure}
\label{subsec:RG procedure}

We begin by describing our RG procedure, which will allow us to
determine the shear viscosity of interacting graphene in the
collisionless regime.  The small parameter justifying our calculation
is the renormalized coupling constant at frequency $\omega$,
$\alpha(\omega)$, which is small at $\omega \ll D$, where $D$ is the
bandwidth of graphene (unlike the bare coupling constant, which is not
small, with $\alpha_0=e^2/(\hbar v \bar{\epsilon}) \approx 2.2$ for
the vacuum case $\bar{\epsilon}=1$).

To obtain the RG equations for the coupling parameter and for the
shear viscosity we perform a leading order RG analysis, which shows
that the Fermi velocity of graphene is renormalized by the Coulomb
interaction and diverges logarithmically with growing RG flow $b=e^l$
where $l$ is the RG flow parameter~\cite{Sheehy2007,Elias2011}:
\begin{equation}
 v \to v \left(1+ \frac{\alpha}{4} \ln b\right).
\end{equation}
This leads to the fact that the flow equation of the coupling constant
is given by
\begin{equation}
 \frac{d \alpha(b)}{d \ln b}
 =
 -\frac{1}{4} \alpha(b)^2.
\end{equation}
which is solved by the following expression for the coupling constant
\begin{equation}
 \alpha(b)
 =
 \frac{\alpha_0}{1+\frac{\alpha_0}{4} \ln(D/b)}
 \:.
 \label{eq:renorm-alpha}
\end{equation}
Simultaneously, the frequency is renormalized by the scaling factor
$Z_{\omega}(b)$ which has the form
\begin{equation}
 Z_{\omega}(b)= \left[1+\frac{\alpha}{4} \ln b \right]b^{-1}.
\end{equation}

Next we consider the behavior of the viscosity tensor under the RG
flow. In distinction to the electrical conductivity, which is scale
invariant in two dimensions, the viscosity has a finite scaling
dimension which is given by the dimensionality $d$ of the system, a
result that follows from momentum conservation and the isotropy of
space\cite{Gian}.  Thus we have the rigorous relation
\begin{equation}
\eta_{\alpha\beta\gamma\delta}\left(\omega,\alpha_{0}\right)
=
b^{-d}\eta_{\alpha\beta\gamma\delta}\left(Z_{\omega}\left(b\right)^{-1}\omega,\alpha\left(b\right)\right).
\label{eq:scaling viscosity}
\end{equation}
The physical viscosity at the bare value $\alpha_{0}$ of the coupling
constant can be expressed in terms of the viscosity at a higher
frequency and a weaker coupling constant, since
$\alpha\left(b>1\right)<\alpha_{0}$.  Scaling is expected to stop at
the scale $b^{*}$ where the renormalized frequency equals the band
width: $\omega/Z_{\omega}\left(b^{*}\right)=D$.  This leads to
\begin{equation}
b^{*}\left(\omega\right)=\frac{D}{\omega}\left(1+\frac{\alpha_{0}}{4}\ln\frac{D}{\omega}\right)
\:,
\end{equation}
 and
 $\alpha\left(\omega\right)=\alpha\left(b^{*}\left(\omega\right)\right)$
 given in Eq.~\eqref{eq:scaling viscosity}. If we insert this result
 into Eq.~\eqref{eq:scaling viscosity} we obtain
\begin{equation}
\eta_{\alpha\beta\gamma\delta}\left(\omega,\alpha_{0}\right)
=
\frac{\omega^{2}\eta_{\alpha\beta\gamma\delta}\left(D,\alpha\left(\omega\right)\right)}{D^{2}\left(1+\frac{\alpha_{0}}{4}\ln\frac{D}{\omega}\right)^{2}}.
\label{eq:RG_visc}
\end{equation}
Our remaining task is to determine the high-frequency viscosity at
weak coupling. For $\eta_{\alpha\beta\gamma\delta}\left(D,\alpha\left(\omega\right)\right)$
we can then perform a perturbation theory calculation to obtain
\begin{equation}
\eta_{\alpha\beta\gamma\delta}\left(D,\alpha\left(\omega\right)\right)=\eta_{\alpha\beta\gamma\delta}^{\left(0\right)}\left(D\right)\left[1+C_{\eta}\alpha\left(\omega\right)+\cdots\right],
\end{equation}
where $\eta_{\alpha\beta\gamma\delta}^{\left(0\right)}\left(D\right)$
is the viscosity of non-interacting electrons at frequency $\omega=D$
and $C_{\eta}$ a numerical coefficient of order unity that we will
determine in the next section.

Equation~(\ref{eq:scaling viscosity}) implies that electron-electron
interactions impact graphene's viscosity in two ways: Firstly, by the
overall prefactor $b_*^{-d} = b_*^{-2}$.  As we shall see, this yields
the frequency-dependent Fermi velocity factor in
Eq.~(\ref{eq:dynviscos}).  Secondly, interactions enter via the
renormalized viscosity $\eta_{\alpha \beta \gamma
  \delta}(D,\alpha(b^*))$ on the right side of Eq.~(\ref{eq:scaling
  viscosity}).  Much of our subsequent calculations will involve
computing this to leading perturbative order in $\alpha(b^*)=
\alpha(\omega)$.  This will require computing the correlation function
$C_{\alpha\beta \gamma\delta}$ within perturbation theory.  Below, we
call the zeroth order, e.g., $\curO(\alpha^0)$, and first order,
$\curO(\alpha^1)$, contributions to this quantity as $C_{\alpha\beta
  \gamma\delta}^{(0)}$ and $C_{\alpha\beta \gamma\delta}^{(1)}$,
respectively.

Diagrammatically, these contributions are depicted in
Figs.~\ref{fig:stress} and \ref{fig:Feynman}. The response of free
Dirac fermions, $C_{\alpha\beta \gamma\delta}^{(0)}$, is shown in
Fig.~\ref{fig:Feynman}, panel (a).  Notably, this contribution has
nothing to do with dissipation (which is absent in any non-interacting
system), but rather describes the nonlocal energy-flow response of
Dirac fermions to an external time-dependent perturbation.  The
correlation function $C_{\alpha\beta\gamma\delta}^{(0)}$ is evaluated
in Section~\ref{subsec:the case without Coulomb interaction}.

Obtaining the perturbative contribution to the dynamical shear
viscosity, $C_{\alpha\beta\gamma\delta}^{(1)}$, requires computing the
leading-order Feynman diagrams in the interaction parameter, as shown
in Fig.~\ref{fig:Feynman}, panels (b)-(e). The corresponding
calculation is presented in Section~\ref{subsec: the viscosity of
  interacting graphene}.

\subsection{Free Dirac fermions}
\label{subsec:the case without Coulomb interaction}

\begin{figure}
 \includegraphics[width=0.8\linewidth]{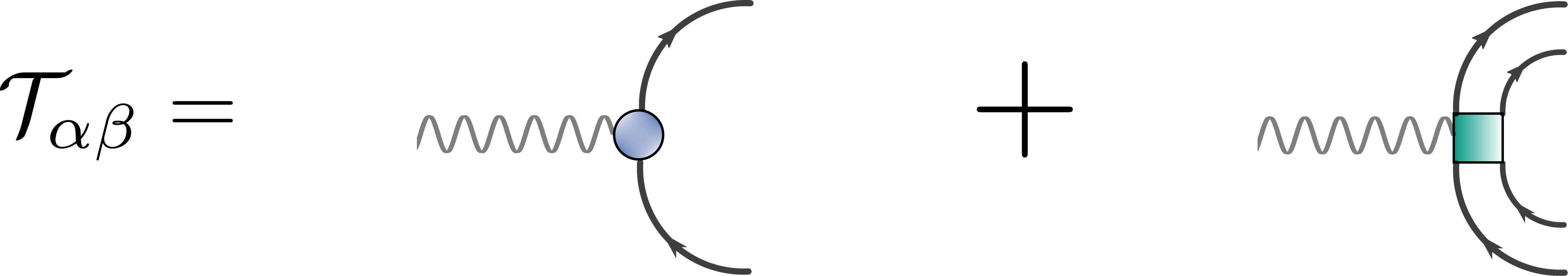}
 \caption{(Color online) Diagrammatic representation of the stress
   tensor vertex. The left vertex describes the non-interacting stress tensor
   (\ref{t0vert}). The right vertex describes the interaction part of
   the stress tensor (\ref{tintvert}).}
 \label{fig:stress}
\end{figure}

\begin{figure}
 \includegraphics[width=0.9\linewidth]{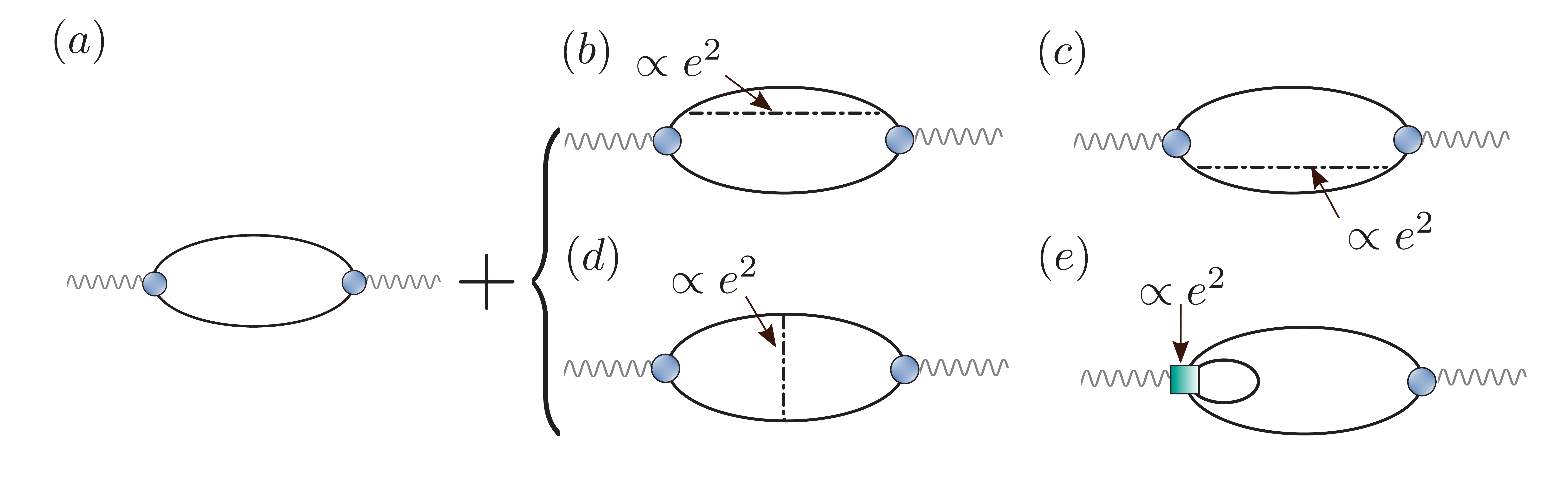}
 \caption{ (Color online) Feynman diagrams for the correlation
   function of the stress tensor. The diagram (a) describes the
   dynamic viscosity of non-interacting Dirac fermions, while the
   remaining diagrams determine the leading order correction due to
   the Coulomb interaction. in the main text the diagrams (b) and (c)
   are referred to as the self-energy diagrams, diagram (d) -- the
   vertex diagram, and diagram (e) as the ``honey diagram'' that
   includes the interaction correction to the stress tensor.}
 \label{fig:Feynman}
\end{figure}

We begin with the zeroth order calculation, which corresponds to the
collisionless dynamic viscosity of noninteracting graphene.  The
Matsubara stress-stress correlation function of a system of
non-interacting Dirac fermions is given by \cite{Principi2016}
\begin{eqnarray}
 \label{eq:correl nonint}
  &&
 C_{\alpha \beta\gamma \delta}^{(0)}(i\Omega)
 \\
 &&
 \nonumber\\
 &&
 \qquad
 =
 T \sum \limits_{\omega} \!\int\frac{d^2k}{(2\pi)^2}\!
 \Tr\! \left[G_{\bs{k},i\omega} {\cal T}_{\alpha \beta}^{(0)}(\bs{k})
   G_{\bs{k},i(\omega+\Omega)} {\cal T}_{\gamma \delta}^{(0)}(\bs{k}) \right]\!,
 \nonumber
\end{eqnarray}
where ${\cal T}_{\alpha \beta}^{(0)}$ are the vertex operators
corresponding to the stress tensor \eqref{eq:contin2}, see
Fig.~\ref{fig:stress}, and $G_{\bs{k}, i\omega}$ are the Matsubara
Green's functions
\begin{equation}
 G_{\bs{k}, i\omega}
 =
 -\frac{(\ii \omega \sigma_0 \!+\! v \bs{k} \!\cdot\! \bs{\sigma})}{(\omega^2 \!+\! (v k)^2)}
 =
 \frac{1}{2}\sum\limits_{s=\pm 1}\frac{\sigma_0\!+\!sv\bs{\sigma}\!\cdot\!\bs{k}/k}{i\omega-svk},
\end{equation}
with $s$ being the band index. The corresponding diagram is shown in
Fig.~\ref{fig:Feynman}, panel (a).


Given that the strain generator (\ref{eq:stress generator J}) is a
combination of the orbital and pseudo-spin parts, we evaluate the
two corresponding contributions to the stress tensor separately. The
orbital contribution is given by
\begin{equation}
 T_{\alpha \beta}^{\mathcal{L},(0)}
 =
 -i [\mathcal{L}_{\alpha \beta}, H_{0}]
 =
 \int\frac{d^2k}{(2\pi)^2} \psi^{\dagger}_{\bs{k},t} \sigma_\alpha k_\beta \psi_{\bs{k},t},
 \label{eq:energystressL}
\end{equation}
where $\psi_{\bs{k},t}$ is the spinor (\ref{spinor}) in momentum space
and time domain. Note that this expression is explicitly not
symmetric. However, adding the pseudo-spin contribution
\begin{equation}
 T^{\mathcal{S},(0)}_{\alpha \beta}
 \!=\!
 -i [\mathcal{S}_{\alpha \beta}, H_{0}]
 =
 \frac{1}{2}\!\int\!\!\frac{d^2k}{(2\pi)^2}\psi^{\dagger}_{\bs{k},t}
 (\sigma_\beta k_{\alpha}\!-\! \sigma_{\alpha} k_{\beta}) \psi_{\bs{k},t},
\end{equation}
where $H_{0}$ is defined by Eq.~\eqref{eq:H0 graphene},
we arrive at the symmetric expression of the stress tensor
\begin{equation}
  \label{tsym}
 T^{(0)}_{\alpha \beta}
 \!=\!
 -i [\mathcal{J}_{\alpha \beta}, H_{0}]
 =\frac{1}{2}\!\int\!\!\frac{d^2k}{(2\pi)^2}
 \psi^{\dagger}_{\bs{k},t} (k_\alpha \sigma_\beta \!+\! k_\beta \sigma_\alpha) \psi_{\bs{k},t}.
\end{equation}
The corresponding vertices (see Fig.~\ref{fig:stress})
\begin{equation}
  \label{t0vert}
  {\cal T}_{\alpha \beta}^{(0)} = \frac{1}{2}(k_\alpha \sigma_\beta \!+\! k_\beta \sigma_\alpha),
\end{equation}
are time-independent and we may perform the sum over the Matsubara
frequencies in Eq.~(\ref{eq:correl nonint}) to obtain:
\[
T\sum\limits_{\omega} \frac{1}{i \omega \!-\! s_1 v k}
\frac{1}{i \omega \!+\!i \Omega \!-\! s_2 v k}
=
\frac{n_F(s_1 v k)\!-\! n_{F}(s_2 v k)}{(s_1\!-\!s_2 ) v k \!+\! i\Omega},
\]
 where $n_F(\omega) = \frac{1}{{\rm e}^{\omega/T}+1}$ is the Fermi
function.

After analytical continuation, $i\Omega\rightarrow\omega+i\delta$, the
imaginary part of this expression is given by a $\delta$-function,
\[
  {\rm Im} \frac{1}{(s_1\!-\!s_2) v k\!+\!\omega \!+\! i \delta}
  = -i \pi \delta(\omega\!-\! (s_1\!-\!s_2) v k),
\]
reflecting the expected behavior that only interband transitions, $s_1\neq s_2$, contribute to
the dynamic viscosity (with $\omega>0$).

The remaining integration is straightforward. As a result, we obtain
for the dynamical shear viscosity of non-interacting Dirac fermions
in pure graphene at charge neutrality:
\begin{equation}
 \eta^{(0)}(\omega)
 =
 \frac{\hbar}{64 v^2} \omega^2,
 \label{eq:visc-nonint}
\end{equation}
which corresponds to the shear viscosity $\eta(\omega)$ of
the standard expression (\ref{eta}) 
for the viscosity tensor $\eta_{\alpha \beta \gamma \delta}$.

The above calculation demonstrates the importance of the pseudo-spin
structure of the fermionic excitations in multi-band systems (in other
words, of the interband transitions). Evaluating the viscosity tensor
using the orbital part of the stress-tensor (\ref{eq:energystressL})
only, one arrives at $\eta_{\alpha \beta \gamma \delta}$ with the
tensor structure that explicitly violates Eq.~(\ref{eta}). The reason
for this incorrect result is that Eq.~\eqref{contin} is not fulfilled.
The fact that physically correct results correspond to the symmetric
stress tensor (\ref{tsym}) is well recognized in literature
\cite{dau2,book,Principi2016} based on the known result (\ref{eta})
for rotationally invariant systems. The problem becomes more difficult
in the anisotropic case \cite{Link2018}, where one does not have the
guidance of the known result. Our derivation of the strain generators
and symmetric stress tensor allows us to establish the structure of
the viscosity tensor from first principles without relying on any
phenomenological assumptions.

\subsection{Dynamic viscosity of interacting graphene}
\label{subsec: the viscosity of interacting graphene}

We now consider the Coulomb interaction correction to the viscosity
correlation function, which we denote as
$C_{\alpha\beta\gamma\delta}^{(1)}$.  Drawing on past experience of
the calculation of the optical conductivity
\cite{Mishchenko2008,Sheehy2009,Teber2014,Link2016} which has shown
the resulting diagrams to be separately divergent, we modify the
Coulomb interaction to
\begin{equation}
  \label{cm}
  V(\bs{r}, \bs{r}')=\frac{e^2 r_0^{-\delta}}{|\bs{r}-\bs{r}'|^{1-\delta}}\:,
\end{equation}
and take the limit $\delta \to 0$ at the end. Here, $r_0$ is a length 
scale introduced to preserve the units of the system at finite $\delta$. 
In the case of the optical conductivity calculation, 
this modification provides a ``soft cut-off'' regularization of the
logarithmically divergent diagrams, such that the divergent
contributions of the self-energy and vertex diagrams cancel out
yielding the finite result (\ref{eq:conductivity}). The Fourier
transform of the modified Coulomb potential (\ref{cm}) is given by
\begin{equation}
 V(\bs{q})=
 \frac{2 \pi \alpha_{\delta}}{|\bs{q}|^{1+\delta}},
 \label{eq:modified-Coulomb}
\end{equation}
where $\alpha_{\delta}=\alpha\: r_0^{-\delta} 2^{\delta} \Gamma[(1\!+\!\delta)/2]/\Gamma[(1\!-\!\delta)/2]$.
We note that the reason for the introduction of the exponent $\delta$
instead of using a screened Coulomb potential is that we are
evaluating the shear viscosity at the charge neutrality point and in
the collisionless regime, $\omega \tau_{ee}\gg1$, which leads to the
fact that the calculation can be performed at $T=0$. Hence, there is
no charge density to screen the Coulomb potential.

In addition, the validity of the present regularization scheme is
ensured by the fact that it reproduces the same value for the
coefficient $\mathcal{C}_{\sigma}$ in the optical conductivity
\eqref{eq:conductivity} found in the Dirac model and the tight-binding
model of graphene~\cite{Link2016}.

In distinction to the charge current, the momentum current of our system 
contains single-particle contributions, Eq.~\eqref{tsym}, and many-body 
contributions. The latter take into account the flow of momentum by
interaction effects.
Our first task is to  evaluate this \emph{interaction part of the
  stress tensor}
\begin{equation}
  \label{tint}
  T_{\alpha \beta}^{(\text{int})}(\bs{r}) = -i[ \mathcal{J}_{\alpha \beta},H_{\text{int}}].
\end{equation}
\begin{widetext}
Explicitly, we find\cite{Schwinger1959}
\begin{eqnarray}
&&
T_{\alpha \beta}^{(\text{int})}(\bs{r})
=
\!\int\! d^2r_1d^2r_2
\psi^{\dagger}({\bs{r_1}})\psi^{\dagger}({\bs{r_2}})\psi({\bs{r_2}})\psi({\bs{r_1}})
(r_{1\alpha}-r_{2\alpha}) \nabla_{\bs{r_2}\beta} V(\bs{r_1},\bs{r_2})
\\
&&
\nonumber\\
&&
\qquad\qquad\qquad
=
\frac{(1\!-\!\delta)r_0^{-\delta} }{2}  
\!\int\! d^2r_1d^2r_2
\psi^{\dagger}({\bs{r_1}})\psi^{\dagger}({\bs{r_2}})\psi({\bs{r_2}})\psi({\bs{r_1}})
\frac{(r_{1\alpha}-r_{2\alpha})(r_{1\beta}-r_{2\beta})}{|\bs{r_1}-\bs{r_2}|^{3-\delta}}.
\nonumber
\end{eqnarray}
The integrated stress-tensor can be obtained by calculating the
zero-momentum Fourier component of $T_{\alpha \beta}^{(\text{int})}(\bs{r})$,
\begin{equation}
  \label{tintvert}
T_{\alpha \beta}^{(\text{int})}(\bs{q}=\bs{0},\tau)
=
\frac{1-\delta}{2} r_0^{-\delta} 2^{1-\delta} \pi
\frac{\Gamma[(3\!+\!\delta)/2]}{\Gamma[(3\!-\!\delta)/2]}
\int\frac{d^2l}{(2\pi)^2}
\frac{l_\alpha l_\beta}{|\bs{l}|^{3+\delta}} n(\bs{l},\tau) n(-\bs{l},\tau),
\end{equation}
\end{widetext}
with the density operators in momentum space,
\begin{equation}
 n(\bs{l}, \tau)
 =
 \int\frac{d^2p}{(2\pi)^2}
  \psi^{\dagger}_{\bs{p}}(\tau) \psi_{\bs{p+l}}(\tau) 
  \:.
\end{equation}

The remaining calculation of the interaction correction to the dynamic
viscosity amounts to the evaluation of the four Feynman diagrams shown in
Fig.~\ref{fig:Feynman}, panels (b)-(e). In contrast to the similar
calculation of the optical conductivity
\cite{Mishchenko2008,Sheehy2009,Teber2014,Link2016}, these include an
additional diagram, see Fig.~\ref{fig:Feynman}, panel (e), describing
the correlation between the interaction part of the stress tensor
$T_{\alpha \beta}^{(\text{int})}$ and the non-interacting part
$T_{\alpha\beta}^{(0)}$. In what follows we will refer to this diagram
as the ``honey diagram'', since the high viscosity of classical fluids 
such as honey arises
mostly due to the strong interaction between the fluid molecules
leading to a large contribution of $T_{\alpha\beta}^{(\text{int})}$.

Computing the contributions of different diagrams separately, we
arrive at the interaction correction to the correlation function
 in the form
\begin{equation}
  C_{\alpha\beta\gamma\delta}^{(1)}=C_{\alpha\beta\gamma\delta}^{(1,bc)}+C_{\alpha\beta\gamma\delta}^{(1,d)}
  +C_{\alpha\beta\gamma\delta}^{(1,e)}.
\end{equation}
Given the tensor structure (\ref{eta}) it is sufficient to evaluate
just one component of the viscosity tensor. Focusing on $C_{xyxy}^{(1)}$,
we write the three different contributions to the correlation function
as
\begin{widetext}
\begin{equation}
C_{xyxy}^{(1,bc)}(i\Omega) 
= -2 N \int \frac{\omega}{2 \pi} \int\frac{d^2p}{(2\pi)^2}
{\rm Tr}\left[ G_{\bs{p}, i(\omega+\Omega)} {\cal T}_{xy}^{(0)}(\bs{p}) G_{\bs{p}, i\omega}
  \Sigma(\bs{p}) G_{\bs{p}, i\omega} {\cal T}_{xy}^{(0)}(\bs{p}) \right],
\label{eq:SELF}
\end{equation}
with the self-energy
\begin{equation}
\Sigma(\bs{p})=
\int \frac{d \omega'}{2 \pi} \int \frac{d^2 k}{(2 \pi)^2} \frac{2 \pi \alpha_{\delta}}{|\boldsymbol{p}-\boldsymbol{k}|^{1+\delta}} G_{\bs{p},i\omega'}
= \phi(\bs{p}) \bs{p}\!\cdot\!\bs{\sigma},
\qquad
\phi(\bs{p})=
\alpha r_0^{-\delta} \frac{2^\delta \Gamma(\frac{\delta}{2})}{8 \Gamma(\frac{4-\delta}{2})} p^{-\delta},
\end{equation}
\begin{equation}
\label{eq:VERTEX}
C_{xyxy}^{(1,d)}(i\Omega) 
= 
N \alpha_{\delta} \int \frac{d\omega d \omega'}{(2 \pi)^2}\int\frac{d^2p}{(2\pi)^2}\frac{d^2k}{(2\pi)^2}
\frac{2 \pi}{|\bs{p}-\bs{k}|^{1+\delta}}
     {\rm Tr} \left[G_{\bs{p}, i\omega} {\cal T}_{xy}^{(0)}(\bs{p}) G_{\bs{p}, i(\omega+\Omega)} G_{\bs{k}, i(\omega'+\Omega)}
       {\cal T}_{xy}^{(0)}(\bs{k}) G_{\bs{k}, i\omega'} \right],
\end{equation}
\begin{equation}
C_{xyxy}^{(1,e)} (i\Omega) =4 N \int \frac{d \omega}{(2 \pi)^2}\int\frac{d^2k}{(2\pi)^2}
{\rm Tr} \left[ G_{\bs{k}, i\omega} {\cal T}_{xy}^{(\mathrm{int})}(\bs{k}) G_{\bs{k}, i(\omega+\Omega)}
  {\cal T}_{xy}^{(0)}(\bs{k}) \right].
\end{equation}
\end{widetext}

\emph{Self-energy and vertex diagrams.} 
The dynamic viscosity (\ref{eq:arriveat}) is determined by the
imaginary part of $C_{\alpha\beta\gamma\delta}^{(1)}$ (after
analytical continuation to real frequencies, $i\Omega\to \omega +
i\delta$). In the self-energy and vertex contributions,
Eqs.~(\ref{eq:SELF}) and (\ref{eq:VERTEX}), the imaginary part is
``less divergent'' than the real part. This can be seen by using the
Kramers-Kronig relations to analyze the Matsubara frequency dependence
of the bare bubble, which has the form $C_{\alpha \beta \gamma \delta}
\propto a \Lambda^2 + b \Lambda \Omega^2+ i \mathcal{N} \Omega^3$
where $\Lambda$ is an ultraviolet cutoff (proportional to the
bandwidth) and $a$, $b$, and $\mathcal{N}$ are numerical
coefficients. The imaginary part is determined by the third term of
this expression and is free of any ultra-violet divergences.
Anticipating that $C_{xyxy}^{(1,bc)}(i\Omega)$ and
$C_{xyxy}^{(1,d)}(i\Omega)$ will have the same form, we proceed by
carefully subtracting the first two contributions in these expressions
that are proportional to $\Lambda^2$ and $\Lambda$. For the
functions analytically continued to the real frequency axis, we find 
for the self-energy diagram

\begin{eqnarray}
  &&
  \im~ C^{(1,bc)}_{xyxy}/\omega
  = -\alpha  \frac{4^{\delta -5} (\delta-4)\omega^{-\delta} r_0^{-\delta}\Gamma(\delta/2)}
  {\cos(\pi\delta/2)\Gamma(2-\delta/2)} \omega^2
  \nonumber\\
 && 
 =
 \left[\frac{1}{2\delta} + \mathcal{N}_{self}-\frac{ \ln(r_0 \omega/4)}{2}+ \mathcal{O}(\delta)\right]
 \alpha(\omega) \eta^{(0)}(\omega) ,
\end{eqnarray}
with $\mathcal{N}_{self}=1/8-\gamma/2$ and for the vertex diagram 
\begin{eqnarray}
  &&
  \im~ C^{(1,d)}_{xyxy}/\omega 
  \\
  &&
  =
  \left[ -\frac{1}{4 \delta} + \mathcal{N}_{vertex} + \frac{\ln(r_0 \omega/4)}{4}+ \mathcal{O}(\delta) \right]
  \alpha(\omega) \eta^{(0)}(\omega), \nonumber
\end{eqnarray}
with $\mathcal{N}_{vertex}=-193/80+(\gamma + 2\pi)/4$, where the above 
expressions are valid for small $\delta$. The details 
of the calculation  to determine the numerical coefficients 
$\mathcal{N}_{self}$ and $\mathcal{N}_{vertex}$ can be found in the appendix.

The self-energy and vertex diagrams, which describe
single-particle propagation, are still divergent for $\delta\to0$. In
contrast to the case of the optical conductivity (with similar
diagrams), these divergences do not cancel when summed, indicating
that the final ``honey'' diagram (panel e of Fig.~\ref{fig:Feynman})
must contribute additional divergent contributions.

\emph{Honey diagram.} The above divergence of the self-energy and
vertex diagrams at $\delta\to0$ will be canceled if the remaining
``honey'' diagram has the form
\begin{eqnarray}
 \label{eq:honey-diagram}
  &&\im~ C^{(1,e)}_{xyxy}/\omega \\
  &&=\left[-\frac{1}{4 \delta} + \mathcal{N}_{honey} 
  + \frac{\ln(r_0 \omega/4)}{4}+ \mathcal{O}(\delta) \right]
  \alpha(\omega) ~\eta^{(0)}(\omega). \nonumber
\end{eqnarray}
To show that this is indeed the case we begin with the Matsubara 
expression of the honey diagram, which contains one
noninteracting stress tensor vertex and one vertex from the
interacting part of the stress tensor:
\begin{widetext}
\begin{eqnarray}
  C_{xyxy}^{(1,e)} (i\Omega) =
  -2\pi(1\!-\!\delta) r_0^{-\delta} 2^{3+\delta}
  \frac{\Gamma[(3+\delta)/2]}{\Gamma[(3-\delta)/2]}
 T^2\sum_{\omega\omega'}\int\frac{d^2l}{(2\pi)^2}\frac{d^2k}{(2\pi)^2} 
  \frac{l_\alpha l_\beta}{|\bs{l}|^{3+\delta}} 
  {\rm Tr}\left[ G_{\bs{k}+\bs{l}, i\omega'} G_{\bs{k},i(\omega+ \Omega)} {\cal T}_{xy}^{(0)} G_{\bs{k},i\omega}\right].
\end{eqnarray}
Summing over the frequencies, we find
\begin{equation}
C_{xyxy}^{(1,e)} (i\Omega)
 =
  (1\!-\!\delta)r_0^{-\delta} 2^{3+\delta} \pi
 \frac{\Gamma[(3+\delta)/2]}{\Gamma[(3-\delta)/2]}  \frac{1}{(2 \pi)^4}
 \int \limits_0^{\infty}\!\! d k \int \limits_0^{\infty} d l
 \int \limits_0^{2 \pi} d \alpha \int \limits_0^{2 \pi} d \beta 
 \frac{k^3 l^{1-\delta } \sin2\alpha \cos2\beta \sin(\alpha-\beta)}
      {\left(4 k^2+\Omega ^2\right) \sqrt{k^2+2 k l \cos (\alpha -\beta )+l^2}}.
\end{equation}

Now, we analytically continue the obtained function to real
frequencies and evaluate the imaginary part of the correlation
function. Here we make use of the identity
\begin{equation}
  \frac{1}{4 k^2 \!+\! \Omega^2} \to \frac{\cal P}{4 k^2 \!-\! \omega^2}
  + \frac{i\pi}{4 \omega} \delta\left(k-\frac{\omega}{2}\right)+ \frac{i\pi}{4 \omega}
  \delta\left(k+\frac{\omega}{2}\right),
  \label{eq:matsubara analytic con}
\end{equation}
where ${\cal P}$ denotes the principal value. The resulting imaginary
part of the correlation function is given by
\begin{equation}
\frac{\im C_{xyxy}^{(1,e)}}{\omega}
  = \alpha
  \frac{2^{2\delta-8}(\delta\!-\!1)\omega^{-\delta}  r_0^{-\delta} \Gamma(\delta/2)}
       {\Gamma(3-\delta/2)} \omega ^2
       =
\left[-\frac{1}{4 \delta }+\mathcal{N}_{honey}
  +  \frac{\ln (r_0 \omega/4)}{4} + \mathcal{O}(\delta) \right]
\alpha(\omega) \eta^{(0)}(\omega) \:,
\end{equation}
\end{widetext}
with $\mathcal{N}_{honey}=1/16+\gamma/4$,
which has the exact form as Eq.~\eqref{eq:honey-diagram} and 
thus cancels the divergence of the self-energy and vertex
diagrams. As a result, the perturbative expression for the
conductivity, on the right side of Eq.~(\ref{eq:scaling viscosity}), has the form of
Eq.~\eqref{eta} with
\be
\eta(D,\alpha(\omega)) = \eta^{(0)}(D)\big[1+ \mathcal{C}_{\eta}\alpha(\omega) \big]\:,
\ee
and the correction coefficient
\begin{equation}
\mathcal{C}_{\eta}
 =
 \mathcal{N}_{self}+\mathcal{N}_{vertex}+\mathcal{N}_{honey}
 =
 \frac{89\!-\!20\pi}{40}
 \approx
 0.65.
\end{equation}
When we use Eq.~\eqref{eq:visc-nonint} and insert 
this into the RG equation Eq.~(\ref{eq:RG_visc}), we finally
arrive at Eq.~\eqref{eq:dynviscos}.

Thus, in contrast to the case of the optical conductivity of graphene,
the dynamic viscosity of graphene reveals significant interaction
corrections.  These are due to the velocity renormalization but also
due to the interaction correction $\mathcal{C}_{\eta}$ that is much
larger than the corresponding correction $\mathcal{C}_{\sigma}=0.01$
in the optical conductivity.

\section{Connection between viscosity and thermal conductivity}
\label{sec:thermo}

In this Section, we relate graphene's dynamic viscosity to its
nonlocal (momentum dependent) energy flow expressed in terms of the 
thermal conductivity. Our derivation is
similar to one presented by Bradlyn \emph{et al.} \cite{Bradlyn12}
relating the viscosity to the momentum-dependent conductivity tensor
$\sigma_{\nu\beta}(\bs{q},\omega)$ in a Galilean invariant (GI)
system. The relation derived in Ref.~\onlinecite{Bradlyn12} relies on
two facts. Firstly, the continuity equation Eq.~(\ref{Eq:continuity})
allows one to relate correlation functions of the momentum to strain
correlation functions (and, hence, the viscosity tensor). Following
Bradlyn \emph{et al.}, the relation is (with $\omega^+=\omega+i0^+$,
and in the limit of $\bs{q}\to0$):
\bea
\nonumber
&&
(\omega^+)^2\int\limits_0^\infty\!dt\;
    {\rm e}^{i\omega^+ t}\int\! d^2x {\rm e}^{-i\bs{q}\cdot\bs{x}}
\langle [ g_{\nu}(\bs{x},t) , 
 g_{\beta}(0,0)]\rangle
\\
&&
\nonumber\\
&&
\qquad
 = q_\lambda q_\rho \omega^+ 
\left(\eta_{\lambda\nu\rho\beta}(\omega)  
+\frac{i\kappa^{-1}}{\omega^+}\delta_{\lambda\nu}\delta_{\rho\beta}\right),
\label{eq:thermalconnect}
\eea
with $\kappa$ being the compressibility.

The second fact used by Bradlyn {\it et al.} is that, in a system with
GI, the momentum density is proportional to the particle current, so
that the momentum correlation function can be related to a current
correlation function which, within the Kubo formalism, determines the
optical conductivity.  This then leads to the relation
\be
\sigma_{\nu\beta}(\bs{q},\omega) 
=\frac{in\delta_{\nu\beta}}{m\omega^+}+
\frac{q_\lambda q_\rho}{m(\omega^+)^2} \Big(\eta_{\lambda\nu\rho\beta}(\omega) 
+\frac{i\kappa^{-1}}{\omega^+} \delta_{\lambda\nu}\delta_{\rho\beta}\Big),
\label{Eq:sigmaresult}
\ee
connecting the electrical conductivity to the viscosity tensor in a GI
system (equivalent to Eq.(4.9) of Ref.~\onlinecite{Bradlyn12} in the
limit of $B\to 0$). Here, $n$ is the average charge density.

In graphene, the lack of GI implies that the momentum current is not
proportional to the charge current and the relation
(\ref{Eq:sigmaresult}) does not hold. However, since
Eq.~(\ref{eq:thermalconnect}) still holds, it is natural to ask if it
can be used to derive an alternate relation connecting the viscosity
tensor to a response function of graphene.  To do this, we note that
the momentum density Eq.~(\ref{eq:modified g}) is proportional to the
energy current in graphene \cite{Lucas2018,Narozhny2017,Briskot2015}.
We can see this by considering the noninteracting energy density
operator
\be
\varepsilon(\bs{x}) = - \frac{iv}{2}
\left[
\psi^\dagger(\bs{x}) \bs{\sigma}\!\cdot\bs{\nabla}\psi(\bs{x})
\!-\!
\left[\bs{\sigma}\!\cdot\bs{\nabla} \psi^\dagger(\bs{x})\right]\psi(\bs{x})\right],
\ee
which satisfies the continuity equation 
\be
\bs{\nabla}\!\cdot\bs{Q}(\bs{x},t) + \partial_t \varepsilon(\bs{x},t) = 0 ,
\ee
with $\bs{Q}(\bs{x},t) = v^2 \bs{g}(\bs{x},t)$, so that, indeed, the
energy current is directly proportional to the momentum density
Eq.~(\ref{eq:modified g}).

Using this connection along with Eq.~(\ref{eq:thermalconnect}), we now
proceed to relate the nonlocal thermal conductivity to the viscosity
tensor. Following Luttinger \cite{Luttinger1964}, we add a
time-dependent perturbation to our system Hamiltonian,
\[
H_1 = \int\! d^2x \; \varepsilon(\bs{x}) \chi(\bs{x},t),
\quad
\chi(\bs{x},t) = e^{-i\omega^+ t} \chi(\bs{x}),
\]
allowing us to incorporate, e.g. a local temperature gradient,
$\bs{\nabla}\chi(\bs{x})=-\bs{\nabla}T/T$ (with $\omega$ a frequency
scale of the temperature oscillations). Following the standard linear
response theory \cite{Mahan}, we obtain the frequency-dependent heat
current
${\langle{Q}_\alpha\rangle=-\kappa_{\alpha\beta}(\bs{q},\omega)\partial_\beta{T}}$
with the thermal conductivity tensor 
\bea
\kappa_{\alpha\beta}(\bs{q},\omega) &=& \frac{v^4}{iT\omega^+} K_{\alpha\beta}(\bs{q},\omega),
\\
&&
\nonumber\\
K_{\alpha\beta}(\bs{q},\omega)
&\equiv &
-\!\!\int\! dt\; e^{i\omega^+ t} \!\!\int\! d^2x\; e^{-i\bs{q}\cdot\bs{x}}
\langle \left[g_\alpha(\bs{x},t),g_\beta(0,0)\right]\rangle,
\nonumber
\eea
in terms of a Fourier-transformed correlation function
$K_{\alpha\beta}(\bs{q},\omega)$ of the momentum density. Now using
Eq.~(\ref{eq:thermalconnect}) and taking the large $\omega$ limit (in
which we may neglect the term proportional to the inverse
compressibility), we finally arrive at
\be
 \kappa_{\alpha\beta}(\bs{q},\omega) = -\frac{v^4}{T (\omega^+)^2}
q_\lambda q_\rho  \eta_{\lambda\alpha\rho\beta}(\omega)  ,
\ee
the desired relation between the frequency-dependent viscosity tensor
and the nonlocal thermal conductivity.

Dropping the infinitesimal part of the frequency and plugging in our
main result, Eq.~(\ref{eq:dynviscos}), we obtain (assuming the
standard frequency-dependent renormalization of the velocity):
\be
\kappa_{\alpha\beta}(\bs{q},\omega) = -\frac{\hbar v(\omega)^2q^2}{64T}
\delta_{\alpha\beta} \left(1+{\cal C}_{\eta}\alpha\left(\omega\right)\cdots\right),
\ee
for the nonlocal thermal or energy-flow response of graphene in the
collisionless regime.

\section{Conclusions}
\label{conclusions}

We determined the elastic response of graphene in the collisionless
regime and related it to the nonlocal energy flow response of the
system. In doing so we extended the theoretical framework of Bradlyn
\emph{et al.} \cite{Bradlyn12} in which the viscosity was derived
using strain generators to systems with pseudo-spins and showed that
the pseudopsin also contributes to the shear viscosity in the
collisionless regime and cannot be neglected.

In particular, we demonstrated that the Coulomb interaction between
the quasiparticles of graphene has a sizable influence on the shear
viscosity of graphene in the collisionless regime. Therefore, the
self-energy diagram, the vertex diagram and the honey diagram were
evaluated using a soft cut-off on the Coulomb potential. The momentum
flux of the system is then governed by comparable single-particle and
many-particle contributions.  The correction coefficient in first
order of the coupling constant determined out of the sum of these
Feynman diagrams is given by $\mathcal{C}=(89-20 \pi)/40 \approx0.65$.
The influence of this value of the correction coefficient can be seen
in Fig.~\ref{fig:fig1}.

\section{Acknowledgement}

This work
was performed in part at the Aspen Center for Physics, which is
supported by National Science Foundation grant PHY-1607611.
J.M.L. thanks the Carl-Zeiss-Stiftung for financial support. DES also
acknowledges support from NSF grant No. DMR-1151717. BNN acknowledges
support from the MEPhI Academic Excellence Project, Contract
No. 02.a03.21.0005


\appendix

\section{Correction coefficient of the viscosity in the collisionless regime}

Here, we give a detailed presentation of the calculation of the different Feynman diagrams 
contributing to the correction coefficient $\mathcal{C}_{\eta}$. These diagrams are the 
\emph{self-energy diagram}, the \emph{vertex diagram} and the \emph{honey diagram}.
To evaluate the different diagrams, we introduce a soft cut-off to the Coulomb interaction 
$V_\delta(\bs{q})=2 \pi \alpha_{\delta}/|\bs{q}|^{1+\delta}$ where the small parameter $\delta$ regularizes 
the integrals. The coupling constant $\alpha_{\delta}$ is defined as
$
	\alpha_{\delta} 
	= 
	\alpha_0 ~  r_0^{-\delta} \zeta_{\delta}
$
with
$
      \zeta_{\delta}
      =
      \frac{2^{\delta}  \Gamma\left( \frac{1 + \delta}{2} \right)}{\Gamma \left( \frac{1-\delta}{2} \right)}
      \:,
$
where we introduced the length scale $r_0$ in such a way that the dimensionality of Coulomb interaction remains unchanged. 

\subsection{The self-energy diagram}

We start with the evaluation of the self energy which is defined as
\begin{equation}
 \Sigma(\bs{k})
 =
 \!\!\int\!\! \frac{d \omega}{2 \pi} \frac{d^2 p}{(2 \pi)^2}
 \frac{2 \pi \alpha_{\delta}}{|\bs{p}\!-\!\bs{k}|^{1+\delta}} G_{\bs{p},i\omega}
 \!=\!
 \phi(\bs{k}) \bs{k} \!\cdot\! \bs{\sigma} \:,
\end{equation}
with
\begin{equation}
 \phi(\bs{k})=\mathcal{A} k^{-\delta}
 =
 \alpha_0 r_0^{-\delta} \frac{2^\delta \Gamma(\delta/2)}{8 \Gamma(2-\delta/2)} k^{-\delta}.
 \end{equation}
The correlation function of the self-energy diagram is given by

\begin{widetext}
  
  \begin{eqnarray}
    &&
  C_{xyxy}^{(1,bc)}(i\Omega) = -8 \int\limits_P
  \Tr\left[ G_{\bs{p}, i(\omega+\Omega)} {\cal T}_{xy}^{(0)}(\bs{p}) G_{\bs{p},i \omega}
    \Sigma(\bs{p}) G_{\bs{p}, i \omega} {\cal T}_{xy}^{(0)}(\bs{p}) \right]
  \nonumber\\
  &&
  \nonumber\\
  &&
  \qquad\qquad\qquad\qquad\qquad\qquad
 =
 -\alpha_0  r_0^{-\delta} \frac{1}{4}
 \frac{2^\delta \Gamma(\delta/2)}{\Gamma(2-\delta/2)}
 \int \frac{d^2 k}{(2 \pi)^2} \frac{d \omega}{2 \pi}
 \frac{p^{-\delta}}{(k^2\!+\!(\omega\!+\!\Omega)^2)} \frac{\Tr(\mathcal{B})}{(k^2 \!+\! \omega^2)^2},
\end{eqnarray}
with
\begin{equation}
\Tr(\mathcal{B})=
    -2 k_x (k_x+k_y) \big(k_x^4+k_x^2 \big[\omega  (\omega +2 \Omega )-10 k_y^2 \big] 
    +
    k_y^2 \big[5 \left(k_y^2+\omega ^2\right)+2 \omega  \Omega \big]\big).
\end{equation}
After performing the frequency integral and the integration over the
angle, we obtain
\begin{equation}
 C_{xyxy}^{(1,bc)} (i \Omega)
 = 
 \alpha_0 \int \limits_0^\infty d k
 \frac{2^{\delta -4} \Gamma(\delta/2) k^{4-\delta } \left(4 k^2-\Omega ^2\right)}
      {\pi  \Gamma(2-\delta/2) \left(4 k^2+\Omega ^2\right)^2}
 \:.
\end{equation}
In order to determine the numerical coefficient of the imaginary part
of the correlation function $C_{xyxy}^{(1,bc)}(i \Omega)$, we have to
calculate the difference
\begin{equation}
  f_{xyxy}^{(1,bc)}(i \Omega) =  \frac{C_{xyxy}^{(1,bc)} (i \Omega)-C_{xyxy}^{(1,bc)} (0)}{ \Omega^2}
  =
  - \alpha_0 \int \limits_0^\infty d k
  \frac{2^{\delta -6} \Gamma \left(\frac{\delta }{2}\right) k^{2-\delta } \left(12 k^2+\Omega ^2\right)}
       {\pi  \Gamma \left(2-\frac{\delta }{2}\right) \left(4 k^2+\Omega ^2\right)^2}
       \:,
\end{equation}
and
\begin{eqnarray}
&&
  \frac{f_{xyxy}^{(1,bc)}(i \Omega)-f_{xyxy}^{(1,bc)}(0)}{\Omega}
   =  \alpha_0
   \int \limits_0^\infty d k
   \frac{2^{\delta -8} \Omega  \Gamma(\delta/2) k^{-\delta } \left(20 k^2+3 \Omega ^2\right)}
        {\pi  \Gamma(2-\delta/2) \left(4 k^2+\Omega ^2\right)^2}
   = - \alpha_0
   \frac{4^{\delta -5} (\delta -4) \Omega^{-\delta} \Gamma(\delta/2)}
        {  \cos(\pi\delta/2)\Gamma(2-\delta/2)} \:,
     \nonumber\\
  &&
  \nonumber\\
  &&
  \qquad\qquad\qquad\qquad\qquad\qquad
   \approx
   \frac{ \alpha_0}{128 ~\delta }+  \alpha_0
   \frac{-4 \ln (r_0 \Omega )-4 \gamma +1+4 \ln (4)}{512} + \mathcal{O}(\delta),
\end{eqnarray}
where in the last step we expanded the expression for small
$\delta$. The self-energy diagram diverges upon taking the limit
$\delta \to 0$.  The other two Feynman diagrams are going to cancel
this divergence.

\subsection{The vertex diagram}

In this section we focus on the vertex diagram. The vertex diagram is
defined by the following correlation function
\begin{equation}
C_{xyxy}^{(1,d)}(\ii \Omega) = -
	\int \frac{d^2 p}{(2 \pi)^2} \frac{d \omega}{2 \pi}
	\int \frac{d^2 k}{(2 \pi)^2} \frac{d \omega'}{2 \pi}
	\frac{2 \pi \alpha_{\delta}}{|\bs{p}-\bs{k}|^{1+\delta}}  
	\Tr \left[G_{\bs{p},\ii \omega} {\cal T}_{xy}^{(0)}(\bs{p}) G_{\bs{p}, i (\omega+\Omega)}
          G_{\bs{k},i (\omega'+\Omega)} {\cal T}_{xy}^{(0)}(\bs{k}) G_{\bs{k},i \omega'} \right].
\end{equation}
After inserting the corresponding expressions of the Green's functions
and the energy-stress tensor and performing the two frequency
integrals, we find
%
%
\begin{eqnarray}
 &&
	C_{xyxy}^{(1,d)}(i \Omega)  
	= 
	-\int \frac{d^2 p}{(2 \pi)^2} \frac{d^2 q}{(2 \pi)^2} 
	\frac{2 \pi \alpha_{\delta}}{|\bs{p}-\bs{q}|^{1+\delta}} 
	\frac{2}{p q \left(4 p^2+\Omega ^2\right) \left(4 q^2+\Omega ^2\right)}
        \nonumber\\
        &&
        \nonumber\\
        &&
        \qquad\qquad\qquad\qquad
	\times
	\left\{
        p^2 \left[p_x q_x \left(q^2+q_x^2-3 q_y^2\right)+p_y q_y \left(q^2-3 q_x^2+ q_y^2\right)\right]
        \right.
        \nonumber\\
        &&
        \nonumber\\
        &&
        \qquad\qquad\qquad\qquad\qquad
	+
        p_x^3 q_x \left(q^2+q_x^2-3 q_y^2\right)
        +p_x^2 \left[\Omega ^2 \left(q_y^2-q_x^2\right)-3 p_y q_y \left(q^2-3 q_x^2+q_y^2\right)\right]
        \nonumber\\
        &&
        \nonumber\\
        &&
        \qquad\qquad\qquad\qquad\qquad
        \left.
        - 3 p_x p_y^2 q_x \left(q^2+q_x^2-3 q_y^2\right)
        +p_y^2 \left[p_y q_y \left(q^2-3 q_x^2+q_y^2\right)+\Omega ^2 (q_x-q_y) (q_x+q_y)\right] 
	\right\}.
\end{eqnarray}
%
%
Next, we subtract the zero-frequency part from the above expression to obtain 
\begin{equation}
 f_{xyxy}^{(1,d)}(i \Omega)=\frac{C_{xyxy}^{(1,d)}(i \Omega)-C_{xyxy}^{(1,d)}(0)}{\Omega^2}.
\end{equation}
To finally determine the contribution to the correction coefficient,
we have to subtract again the zero-frequency part which yields
\begin{equation}
 \frac{f_{xyxy}^{(1,d)}(i \Omega) - f_{xyxy}^{(1,d)}(0)}{\Omega}=Q_1+Q_2+Q_3,
\end{equation}
where $Q_1$ and $Q_2$ are convergent for $\delta=0$, whereas the
integral $Q_3$ is divergent for $\delta \to 0$.  The explicit
expression of these three integrals are
%
\begin{eqnarray}
 	Q_1
	&=&
	-\frac{\alpha_{\delta}}{2}
	\frac{ \Omega^{-\delta} }{(2 \pi)^2} 
	\int \limits_0^{\infty}  \frac{dp}{p (4 p^2+1)}
	\int \limits_0^{\infty}  \frac{dq}{q (4 q^2+1)}
	\int \limits_0^{\pi} d \varphi 
	\frac{(p q/4) \cos2 \varphi+p^2 q^2 \cos\varphi \cos2 \varphi}
             {\left[ p^2 +q^2 -2 p q \cos\varphi \right]^{(1+\delta)/2}}
             \:,
        \\
        &&
        \nonumber\\
        Q_2
	&=&
	- \alpha_{\delta }
	 \frac{\Omega^{-\delta}}{(2 \pi)^2}
	\int \limits_0^{\infty}  \frac{dp}{ p (4 p^2+1)}
	\int \limits_0^{\infty}  \frac{dq}{ q (4 q^2+1)} 
	\int \limits_0^{\pi} d \varphi
	\frac{p^3 q \cos2 \varphi }{[p^2+q^2-2 pq \cos\varphi]^{(1+\delta)/2}}
	\:,
        \\
        &&
        \nonumber\\
	Q_3
	& = &
	-\frac{ \alpha_{\delta }}{16} \frac{\Omega^{-\delta}}{(2 \pi)^2} 
	\int \limits_0^{\infty}  \frac{p d p}{p^2 (4p^2+1)}
	\int \limits_0^{\infty}  \frac{qd q}{q^2 (4q^2+1)} 
	\int \limits_0^{2 \pi} d \varphi \frac{\cos\varphi\cos2 \varphi
          [ p^2 (4 p^2+1)+ q^2 (4 q^2+1)]}
             {[p^2 +q^2 -2 p q \cos\varphi]^{(1+\delta)/2}} 
             \:.
\end{eqnarray}
In the following, we demonstrate how the different integrals are evaluated.
%
%
%

%
\paragraph{Calculation of $Q_1$}

Since $Q_1$ is convergent for $\delta=0$, we set $\delta=0$, substitute the momentum variable
$q$ by introducing the variable $q = x p$ and obtain the following expression:
\begin{equation}
	Q_1
	=
	-\frac{ \alpha_{\delta }}{2}
	\frac{\Omega^{-\delta}}{(2 \pi)^2} 
	\int \limits_0^{\infty} \frac{dp }{4 p^2+1}
	\int \limits_0^{\infty}  \frac{dx}{4 x^2 p^2+1} 
	\int \limits_0^{\pi} d \varphi
	\frac{(1/4)  \cos2 \varphi+ x p^2 \cos\varphi \cos2 \varphi}
             {\sqrt{ 1 + x^2 -2 x \cos\varphi}}.
\end{equation}
The integration over $p$ can be done easily using
\begin{eqnarray}
	\int \limits_0^{\infty} 
	\frac{dp }{ 4 p^2+1}\frac{1}{ 4 x^2 p^2+1}
	= 
	\frac{\pi}{4(1+x)}, \qquad
	\int \limits_0^{\infty}  
	\frac{dp}{ 4 p^2+1}\frac{p^2}{ 4 x^2 p^2+1}
	= 
	\frac{\pi}{16 x (1+x)}.
\end{eqnarray}
After performing first the $x$-integral and than the angle integral, we find 
\begin{equation}
 Q_1=-\frac{\alpha_{\delta}}{240}
 \:.
\end{equation}
\paragraph{Calculation of $Q_2$}

Here again, we apply the variable substitution $q=x p$ which leads to
the integral
\begin{equation}
	Q_2
	\!= 
	\!- \alpha_{\delta } \frac{\Omega^{-\delta}}{(2 \pi)^2}
	\!\int\limits_0^{\infty}\!  \frac{d p p^{2-\delta}}{4 p^2 \!+\!1}
	\!\int\limits_0^{\infty}\!  \frac{d x}{4 x^2 p^2 \!+\!1} 
	\!\int\limits_0^{\pi}\!  \frac{d \varphi\cos2 \varphi}
             {\left[1\!+\!x^2 \!-\!2 x \cos\varphi \right]^{(1+\delta)/2}} 
	= 
	q \!\int\limits_0^{\infty}\! dx \frac{1\!-\!x^{\delta-1}}{x^2-1}
	\!\int\limits_0^{\pi}\! d \varphi
        \frac{2\cos^2(\varphi) \!-\!1}{\left[1\!+\!x^2 \!-\!2 x \cos\varphi \right]^{(1+\delta)/2}},
\end{equation}
with 
\begin{equation}
	q=-\frac{N \alpha_{\delta }\Omega^{-\delta}}{4(2 \pi)^2}
        \frac{2^{\delta-4} \pi}{\cos(\pi \delta/2)}.
\end{equation}
In order to evaluate this integral, we split up the integral in two
parts. These two integrals are defined as:
\begin{eqnarray}
	Q_{2,a}
	&=&
	2 q \int \limits_0^{\infty} dx \frac{1-x^{\delta-1}}{x^2-1}
	\int \limits_0^{\pi} d \varphi \frac{\cos( \varphi)^2 }{[1+x^2 -2 x \cos\varphi]^{(1+\delta)/2}},
        \\
        &&
        \nonumber\\
	Q_{2,b}
	&=&
	-q \int \limits_0^{\infty} dx \frac{1-x^{\delta-1}}{x^2-1}
	\int \limits_0^{\pi} d \varphi \frac{1}{[1+x^2 -2 x \cos\varphi]^{(1+\delta)/2}}.
\end{eqnarray}

\subparagraph{Analysis of $Q_{2,a}$}

This integral can be split in a singular and a non-singular
contribution $Q_{2,a}= Q_{2,a}^s+Q_{2,a}^{ns}$, where the singular
contribution is given by
\begin{equation}
	Q_{2,a}^s
	=
	2 q \int \limits_0^{\infty} dx \frac{1-x^{\delta-1}}{x^2-1}
	\int \limits_0^{\pi} d \varphi \cos^2\varphi
	= 
	 q \frac{\pi^2}{2}  \cot(\pi\delta/2).
\end{equation}
Next, the non-singular contribution is analyzed, which is defined as
\begin{equation}
	Q_{2,a}^{ns}
	=
	Q_{2,a}-Q_{2,a}^{s} 
	=
	2 q \int \limits_0^{\infty} dx \frac{1-x^{\delta-1}}{x^2-1}
	\int \limits_0^{\pi}d \varphi \cos^2\varphi
	\left[ \frac{1}{ [1+x^2 -2 x \cos\varphi]^{(1+\delta)/2}}-1\right].
\end{equation}
Since this integral is convergent for $\delta=0$, it can be evaluated
in this limit. We find
\begin{equation}
	Q_{2,a}^{ns}=
	2 q \int \limits_0^{\infty} dx \frac{1}{x(1+x)}
	\int \limits_0^{\pi} d \varphi \cos^2 \varphi
	\left[ \frac{1}{\sqrt{1+x^2 -2 x \cos\varphi }}-1\right] 
	=\frac{\pi q}{6} (11-6 \pi +\ln4096),
\end{equation}
where we first performed the $x$-integral and then the $\varphi$-integral.

\subparagraph{Analysis of $Q_{2,b}$}

This integral is again split up into a singular and a non-singular
contribution. The singular contribution is given by
\begin{equation}
	Q_{2,b}^s=
	-q \int \limits_0^{\pi} d \varphi 
	\int \limits_0^{\infty} dx \frac{1-x^{\delta-1}}{x^2-1}
	=-\frac{\pi^2q}{2} \cot(\pi\delta/2).
\end{equation}
The non-singular part, which is convergent and can be evaluated for $\delta=0$, reads
\begin{equation}
	Q_{2,b}^{ns}=Q_{2,b}-Q_{2,b}^{s}
	=-q \int \limits_0^{\infty} dx \frac{1}{x(1+x)} 
	\int \limits_0^{\pi} d \varphi \left[\frac{1}{\sqrt{1+x^2 -2 x \cos\varphi }}  -1 \right] 
	= \frac{\pi q}{2} (\pi-2 \ln4).
\end{equation}
\subparagraph{Result for $Q_2$}

Combining $Q_{2,a}$ and $Q_{2,b}$, we obtain for the integral $Q_2$ 
the following value:
\begin{equation}
	Q_2=Q_{2,a}+Q_{2,b} = Q_{2,a}^{ns}+Q_{2,a}^{s}+Q_{2,b}^{ns}+Q_{2,b}^{s}
	=
	- \frac{\alpha_{\delta}\Omega^{-\delta}}{(2 \pi)^2} \frac{2^{\delta-4} \pi}{\cos(\pi\delta/2)}
          \frac{\pi}{6}(11-3 \pi)
	\approx 
	-\alpha_0 \left( \frac{22}{768} - \frac{2 \pi}{256}\right) ,
\end{equation}
where in the last line we expanded for small $\delta$.

\paragraph{Calculation of $Q_3$}

After the variable substitution $p=x q$, we need to evaluate the
following integral
\begin{equation}
Q_3=-\frac{\alpha_{\delta }\Omega^{-\delta}}{8(2 \pi)^2} 
	\int \limits_0^{\infty}  \frac{d q q^{-\delta}}{ 4q^2+1}
	\int \limits_0^{\infty} d x
        \int \limits_0^{2 \pi} d \varphi\frac{x \cos\varphi\cos2 \varphi }
             {[1 +x^2 -2 x \cos\varphi]^{(1+\delta)/2}},
\end{equation}
which is divergent for $\delta \to 0$.  In the following step we use
the identity:
\begin{equation}
	|\bs{k}|^{-(1+\delta)}
	=
	\frac{1}{\Gamma[(1+\delta)/2]}
	\int \limits_0^{\infty} d z \frac{e^{-k^2 z}}{z^{(1-\delta)/2}},
\label{eq:identity1overk}
\end{equation}
and obtain
%
\begin{eqnarray}
  &&
	Q_3=-  \frac{\alpha_{\delta }\Omega^{-\delta}}{(2 \pi)^2} 
        \frac{2^{1+\delta} \pi}{\Gamma[(1+\delta)/2]\cos(\pi\delta/2)}
	\int \limits_0^{\infty} d x 
	\int \limits_0^{2 \pi} d \varphi\cos\varphi\cos2 \varphi
	\int \limits_0^{\infty} d z 
	\frac{x}{z^{(1-\delta)/2}}
	 e^{-\left(1 +x^2 -2 x \cos\varphi \right) z} \:,
        \\
        &&
        \nonumber\\
        &&
        \qquad
        = 
	-  \frac{\alpha_{\delta}\Omega^{-\delta}}{(2 \pi)^2} \frac{2^{1+\delta} \pi }{\cos(\pi\delta/2)}
	\frac{\pi(\delta +2) \Gamma[(1-\delta)/2] \Gamma(\delta/2)}
             {4\Gamma[(1+\delta)/2]\Gamma(3-\delta/2)}
	\approx 
	-\frac{\alpha_0 }{256 \delta }-\frac{\alpha_0  (4 \ln(r_0 \Omega)+4 \gamma -5-8 \ln2)}{1024}
        + \mathcal{O}(\delta),
        \nonumber
\end{eqnarray}
where in the last line we expanded the above expression for small $\delta$.
%
%

\paragraph{Result of the vertex diagram}

Combining all three integrals yields
\begin{eqnarray}
\frac{f_{xyxy}^{(1,d)}(i \Omega) -f_{xyxy}^{(1,d)}(0)}{\Omega}
	= -\frac{\alpha_0}{256~ \delta}	
	+\frac{\alpha_0  (20   \ln (\Omega  r_0)+40 \pi   +20 \gamma   -193  -40   \ln 2)}{5120 }.	
\end{eqnarray}
The vertex diagram is also divergent for $\delta \to 0$, but does not
fully cancel the divergence of the self-energy diagram. A third
Feynman diagram is needed to cancel all divergences.
\subsection{The honey diagram}
\begin{figure}[ht]
 \centering \includegraphics[width=.4\linewidth]{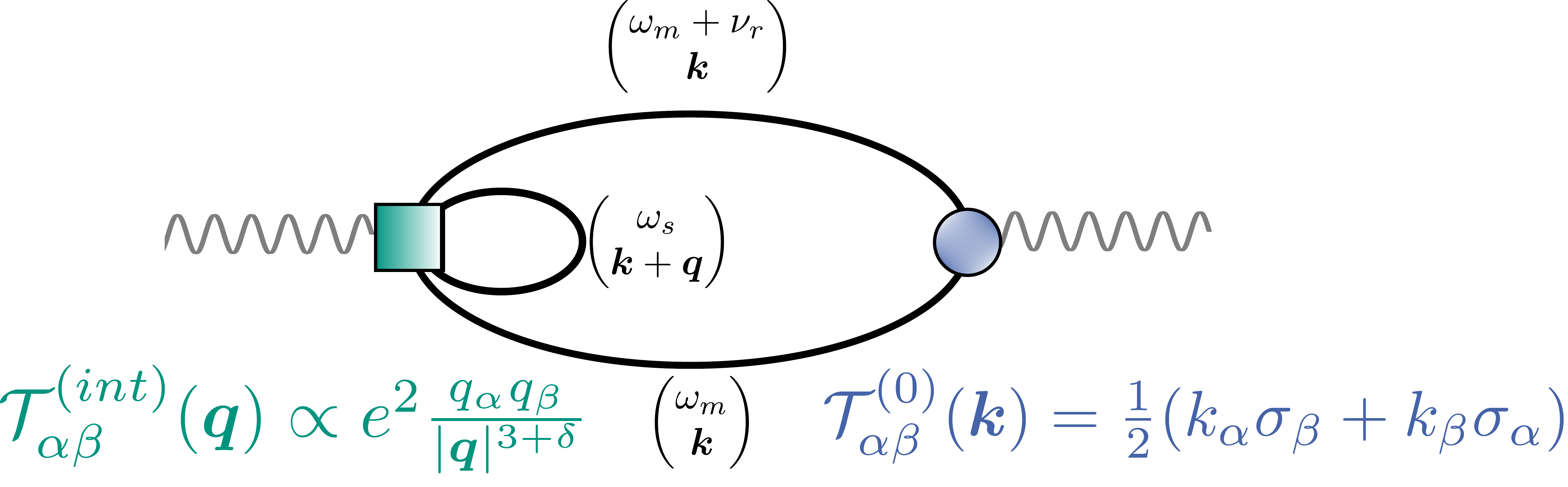}
 \caption[The honey diagram.]{The honey diagram.}
 \label{fig:honeydiagram}
\end{figure}
The last diagram contributing to the correction coefficient
$\mathcal{C}$ is the \emph{honey diagram}, which takes account of 
the interacting part of the energy-stress tensor.  The correlation
function is given by
\begin{equation}
 C_{xyxy}^{(1,e)} (\ii \Omega)=
 - (1-\delta ) r_0^{-\delta} 2^{4+\delta} \pi \frac{\Gamma[(3+\delta)/2]}
 {\Gamma[(3-\delta)/2]}
 \int \limits_{\bs{k},\bs{ l},m,s} 
  \frac{l_\alpha l_\beta}{|\bs{l}|^{3+\delta}} 
 \Tr\left[ G_{\bs{k}+\bs{l},i \omega_s} G_{\bs{k}, i (\omega_m+ \nu_r)} {\cal T}_{xy}^{(0)}(\bs{k}) G_{\bs{k}, i \omega_m} \right]. 
\end{equation}
This Feynman diagram is depicted in Fig.~\ref{fig:honeydiagram}.  
Inserting the explicit formulas of the Green's function and the stress
tensor into the above equation yields
\begin{equation}
C_{xyxy}^{(1,e)} (i \Omega)=
  - (1-\delta)r_0^{-\delta} 2^{3+\delta} \pi \frac{\Gamma[(3+\delta)/2]}
 {\Gamma[(3-\delta)/2]}
 \int\limits_{\bs{k}, \bs{l},m,s} 
 \frac{l_\alpha l_\beta}{|\bs{l}|^{3+\delta}}
 \frac{\Tr[\mathcal{A}]}{(\omega'^2 +|\bs{l}+\bs{k}|^2) (k^2 +\omega^2) [k^2 +(\omega+\Omega)^2]},
\end{equation}
where the trace is given by
\begin{equation}
\Tr[\mathcal{A}]=
-2 \left\{(k_x +l_x) k_y \left[-3 k_x^2+k_y^2+\omega  (\omega +\Omega )\right]
  +(k_y +l_y) k_x \left[k_x^2-3 k_y^2+\omega  (\omega +\Omega )\right]
  +2 k_x k_y \omega' (2 \omega +\Omega )\right\}.
\end{equation}
Next, we perform the two frequency integrations and find
\begin{equation}
C_{xyxy}^{(1,e)}(\ii \Omega)
=
(1-\delta)r_0^{-\delta}  \frac{2^{3+\delta} \pi}{(2 \pi)^4}
\frac{\Gamma[(3+\delta)/2]}{\Gamma[(3-\delta)/2]}  
\int \limits_0^{\infty} d k \int \limits_0^{\infty} d l
\int \limits_0^{2 \pi} d \alpha \int \limits_0^{2 \pi} d \beta 
\frac{k^3 l^{1-\delta } \sin2 \alpha \cos2 \beta  \sin (\alpha -\beta )}
     {\left(4 k^2+\Omega ^2\right) \sqrt{k^2+2 k l \cos (\alpha -\beta )+l^2}}.
\end{equation}
Upon using Eq.~\eqref{eq:matsubara analytic con}, we find that the 
imaginary part of the correlation function can be cast into the form
\begin{equation}
\im C_{xyxy}^{(1,e)} ( \omega)
 =
 (1-\delta)r_0^{-\delta} \frac{2^{3+\delta} \pi}{(2 \pi)^4}
 \frac{\Gamma[(3+\delta)/2]}{\Gamma[(3-\delta)/2]}  
 \int \limits_0^{\infty}  d l \int \limits_0^{2 \pi} d \alpha \int \limits_0^{2 \pi} d \beta 
 \frac{\pi  \omega ^2 l^{1-\delta } \sin2\alpha  \cos2\beta \sin (\alpha -\beta )}
      {16 \sqrt{4 l^2+4 l \omega  \cos (\alpha -\beta )+\omega ^2}}.
\end{equation}
Next, we substitute the angle $\alpha \rightarrow \varphi + \beta$ and perform the $\beta$ integration and obtain
\begin{equation}
\im C_{xyxy}^{(1,e)}(\omega)
=(1-\delta)r_0^{-\delta} \frac{2^{3+\delta} \pi}{(2 \pi)^4}
\frac{\Gamma[(3+\delta)/2]}{\Gamma[(3-\delta)/2]}
\int \limits_0^{\infty}  d l \int \limits_0^{2 \pi} d \varphi 
\frac{\pi^2\omega^2l^{1-\delta }\sin^2\varphi\cos\varphi}{16\sqrt{l^2+l\omega\cos\varphi+\omega^2/4}}.
\end{equation}
%
After using Eq.~\eqref{eq:identity1overk},
we obtain
\begin{equation}
     \im C_{xyxy}^{(1,e)} ( \omega)
 =
   (1-\delta)r_0^{-\delta} \frac{2^{3+\delta} \pi^{5/2}}{(2 \pi)^4}
\frac{\Gamma[(3+\delta)/2]}{\Gamma[(3-\delta)/2]}
  \int \limits_0^{\infty}  d l \int \limits_0^{2 \pi} d \varphi  \int \limits_0^{\infty} d z \,
  \omega ^2 l^{1-\delta } \sin^2\varphi\cos\varphi
  \frac{e^{-\left(l^2+l\omega\cos\varphi+\omega ^2/4\right)}}{16\sqrt{z}}.
\end{equation}
%
As the last step, we first perform the angle-integration over
$\varphi$, than the $l$-integration and at last the $z$-integration,
which leads to
\begin{eqnarray}
 \im C_{xyxy}^{(1,e)} (\omega)=
 \frac{2^{2 \delta -8} (\delta -1) \omega ^{3-\delta}  \Gamma(\delta/2)}{\Gamma(3-\delta/2)}
\approx 
-\frac{\omega ^3}{256 \delta }
+\omega ^3\frac{4\ln(r_0\omega)+4 \gamma+1-8 \ln2}{1024} + \mathcal{O}(\delta),
\end{eqnarray}
where we expanded the above expression for small $\delta$. The honey
diagram is also divergent for $\delta \to 0$. When all three diagrams
are summed up, these divergences cancel each other.
\subsection{The correction coefficient}

Upon combining all three Feynman diagrams, we find for the correction
coefficient $\mathcal{C}_{\eta}$
\begin{equation}
 \mathcal{C}_{\eta}
 =
 \frac{89-20 \pi}{40}
 \simeq
 0.65
 \:.
\end{equation}
This is a relatively large correction coefficient and hence, the
impact of the Coulomb interaction on the shear viscosity of graphene
in the collisionless regime is large.

\end{widetext}


\begin{thebibliography}{10}
\bibitem{Fritz2008}L. Fritz, J. Schmalian, M. Müller \& S. Sachdev,
Phys. Rev. B \textbf{78}, 085416 (2008).

\bibitem{Kashuba2008}A. Kashuba, Phys. Rev. B \textbf{78}, 085415
(2008).

\bibitem{Mueller2009}M. Müller, J. Schmalian \& L. Fritz, Phys. Rev.
Lett. \textbf{103}, 025301 (2009).

\bibitem{dau6}
  L.~D. Landau and E.~M. Lifshitz,
  {\it Fluid Mechanics}
  (Butterworth-Heinemann, Oxford, UK, 2000).

\bibitem{Forster}D. Forster, \emph{Hydrodynamic Fluctuations, Broken
Symmetry, and Correlation Functions}, Frontiers in Physics Vol. 47, (Perseus, New York, 1975), p. 326.

\bibitem{Lucas2018}A. Lucas, K. C. Fong, Journal of Physics: Condensed
Matter \textbf{30}, 053001 (2018).

\bibitem{Narozhny2017}
B. N. Narozhny, I. V. Gornyi, A. D. Mirlin,
and J. Schmalian, Ann. Phys. (Berlin) \textbf{529}, 1700043 (2017).

\bibitem{Titov2013} M. Titov, R. V. Gorbachev, B. N. Narozhny, T.
Tudorovskiy, M. Schütt, P. M. Ostrovsky, I. V. Gornyi, A. D. Mirlin,
M. I. Katsnelson, K. S. Novoselov, A. K. Geim \& L. A. Ponomarenko,
Phys. Rev. Lett. \textbf{111}, 166601 (2013).

\bibitem{Crossno2016} J. Crossno, J. K. Shi, K. Wang, X. Liu, A.
Harzheim, A. Lucas, S. Sachdev, P. Kim, T. Taniguchi, K. Watanabe,
T. A. Ohki \& K. C. Fong, Science \textbf{351}, 1058 (2016).

\bibitem{Ghahari2016} F. Ghahari, H.-Y. Xie, T. Taniguchi, K. Watanabe,
M. S. Foster \& P. Kim, Phys. Rev. Lett. \textbf{116}, 136802 (2016).

\bibitem{Bandurin}D. A. Bandurin, I. Torre, R. Krishna Kumar, M.
Ben Shalom, A. Tomadin, A. Principi, G. H. Auton, E. Khestanova, K.
S. Novoselov, I. V. Grigorieva, L. A. Ponomarenko, A. K. Geim \& M.
Polini, Science \textbf{351}, 1055 (2016).

\bibitem{KrishnaKumar2017} R. Krishna Kumar, D. A. Bandurin, F. M.
D. Pellegrino, Y. Cao, A. Principi, H. Guo, G. H. Auton, M. B. Shalom,
L. A. Ponomarenko, G. Falkovich, I. V. Grigorieva, L. S. Levitov,
M. Polini \& A. K. Geim, Nature Physics \textbf{13}, 1182 (2017).

\bibitem{Briskot2015}U. Briskot, M. Schütt, I. V. Gornyi, M. Titov,
B. N. Narozhny \& A. D. Mirlin, Phys. Rev. B \textbf{92}, 115426 (2015).

\bibitem{Lucas2016}A. Lucas, J. Crossno, K. C. Fong, P. Kim, S. Sachdev,
Phys. Rev. B \textbf{93,} 075426 (2016).

\bibitem{Levitov2016}L. S. Levitov \& G. Falkovich, Nature Phys.
\textbf{12}, 672 (2016).

\bibitem{Seo2017}Y. Seo, G. Song, P. Kim, S. Sachdev, S.-J. Sin,
Phys. Rev. Lett. \textbf{118}, 036601 (2017).

\bibitem{Link2018}J. M. Link, B. N. Narozhny, E. I. Kiselev \& J.
  Schmalian, Phys. Rev. Lett. \textbf{120}, 196801 (2018).

\bibitem{Andreev2011}
  A.V Andreev, S.A. Kivelson, and B. Spivak,
  Phys. Rev. Lett. {\bf 106}, 256804 (2011).

\bibitem{Schuett2011}
  M. Sch\"utt, P.M. Ostrovsky, I.V. Gornyi, and A.D. Mirlin,
  Phys. Rev. B {\bf 83}, 155441 (2011).

\bibitem{Narozhny2015}
  B.N. Narozhny, I.V. Gornyi, M. Titov, M. Sch\"utt, and A.D. Mirlin,
  Phys. Rev. B {\bf 91}, 035414 (2015).

\bibitem{Sheehy2007}D. E. Sheehy, J. Schmalian, Phys. Rev. Lett.
\textbf{99}, 226803 (2007).

\bibitem{Herbut2008} I. F. Herbut, V. Juricic, O. Vafek, Phys. Rev.
Lett. \textbf{100}, 046403 (2008). 

\bibitem{Mishchenko2008}E. G. Mishchenko, Europhys. Lett. \textbf{83},
17005 (2008).

\bibitem{Sheehy2009}D. E. Sheehy, J. Schmalian, Phys. Rev. B \textbf{80},
193411 (2009).

\bibitem{Golub2010}A. Golub and B. Horovitz, Phys. Rev. B \textbf{81},
245424 (2010).

\bibitem{Juricic2010}V. Juricic, O. Vafek, I.F. Herbut, Phys. Rev.
B \textbf{82}, 235402 (2010).

\bibitem{Abedinpour2011}S. H. Abedinpour, G. Vignale, A. Principi,
M. Polini, W.-K. Tse, and A. MacDonald, Phys. Rev. B \textbf{84},
045429 (2011).

\bibitem{Sodemann2012}I. Sodemann and M. M. Fogler, Phys. Rev. B\textbf{
86}, 115408 (2012).

\bibitem{Rosenstein2013}B. Rosenstein, M. Lewkowicz and T. Maniv,
Phys. Rev. Lett. \textbf{110}, 066602 (2013).

\bibitem{Gazzola2013}G. Gazzola, A. L. Cherchiglia, L. A. Cabral,
M. C. Nemes and M. Sampaio, Europhysics Letters \textbf{104}, 27002
(2013).

\bibitem{Teber2014}S. Teber and A. V. Kotikov, Europhysics Letters
\textbf{107}, 57001 (2014).

\bibitem{Link2016}J. M. Link, P. P. Orth, D. E. Sheehy, and J. Schmalian
Phys. Rev. B \textbf{93}, 235447 (2016).

\bibitem{Nair2008}R. R. Nair, P. Blake, A. N. Grigorenko, K. S. Novoselov,
T. J. Booth, T. Stauber, N. M. R. Peres, A. K. Geim, Science \textbf{320},
1308 (2008).

\bibitem{Stauber2008}T. Stauber, N. M. R. Peres, and A. K. Geim,
Phys. Rev. B \textbf{78}, 085432 (2008). 

\bibitem{Conti1999}S. Conti and G. Vignale Phys. Rev. B \textbf{60},
7966 (1999).

\bibitem{Kiselev2018} E. I. Kiselev and J. Schmalian, preprint arXiv:1806.03933 (2018).

\bibitem{Forcella2014}D. Forcella, J. Zaanen, D. Valentinis, and
D. van der Marel, Physical Review B \textbf{90}, 035143 (2014).

\bibitem{Bradlyn12}B. Bradlyn, M. Goldstein, and N. Read, Phys. Rev.
B \textbf{86}, 245309 (2012).

\bibitem{Principi2016}A. Principi, G. Vignale, M. Carrega, and M.
Polini, Phys. Rev. B \textbf{93}, 125410 (2016).

\bibitem{dau4}
  V.B. Berestetskii, E.M. Lifshitz, and L.P. Pitaevskii,
  {\it Quantum Electrodynamics}
  (Pergamon, New York, 1980).

\bibitem{Mecklenburg2011}M. Mecklenburg and B. C. Regan, Phys. Rev.
Lett. \textbf{106}, 116803 (2011); Erratum Phys. Rev. Lett. \textbf{106},
229901(E) (2011).

\bibitem{dau2}
  L.D. Landau and E.M. Lifshitz,
  {\it The Classical Theory of Fields} (Pergamon, New York, 1975).

\bibitem{book}
C. Itzykson and J. Zuber, 
{\it Quantum Field Theory} (McGraw-Hill, New York, 1988).

\bibitem{Belinfante1940}F. J. Belinfante, Physica \textbf{7}, 449
(1940).

\bibitem{Rosenfeld1940}L. Rosenfeld, Acad. R. Belg. Memoirs de
Classes de Science \textbf{18 }, 1536 (1940).

\bibitem{Kubo1957}R. Kubo, J. Phys. Soc. Jpn. \textbf{12}.
570 (1957).

\bibitem{Mahan}
  G.D. Mahan,
  {\it Many Particle Physics}, Plenum, New York, 1990.

\bibitem{Luttinger1964}
J.M. Luttinger, Phys. Rev. {\bf 135}, A1505 (1964).

\bibitem{Gian}
G. A. Inkof, J. Küppers, J. M. Link, B. Goutéraux, and
J. Schmalian, preprint.

\bibitem{Schwinger1959}
P. C. Martin and J. Schwinger,
Phys. Rev. \textbf{115}, 1342 (1959).

\bibitem{Elias2011}D. C. Elias, R. V. Gorbachev,  A. S. Mayorov, 
S. V. Morozov, A. A. Zhukov, P. Blake, L. A. Ponomarenko, 
I. V. Grigorieva, K. S. Novoselov, F. Guinea, and A. K. Geim, 
Nature Physics \textbf{7}, 701 (2011).
\end{thebibliography}
\end{document}